\documentclass[aps,prd,preprint,tightenlines,superscriptaddress,byrevtex]{revtex4}
\usepackage{graphicx} 
\usepackage{amsmath}

\graphicspath{{ps}}

\renewcommand{\arraystretch}{1.1}

\begin{document}


\preprint{\vbox{ \hbox{   }
                 \hbox{BELLE-CONF-0916}
                 \hbox{draft v.4.0}
}}

\title{\boldmath Measurement of the form factors of the decay
  $B^+\to\bar D^{*0}\ell^+\nu_\ell$ and determination of the CKM
  matrix element $|V_{cb}|$}

\affiliation{Budker Institute of Nuclear Physics, Novosibirsk}
\affiliation{Chiba University, Chiba}
\affiliation{University of Cincinnati, Cincinnati, Ohio 45221}
\affiliation{Department of Physics, Fu Jen Catholic University, Taipei}
\affiliation{Justus-Liebig-Universit\"at Gie\ss{}en, Gie\ss{}en}
\affiliation{The Graduate University for Advanced Studies, Hayama}
\affiliation{Gyeongsang National University, Chinju}
\affiliation{Hanyang University, Seoul}
\affiliation{University of Hawaii, Honolulu, Hawaii 96822}
\affiliation{High Energy Accelerator Research Organization (KEK), Tsukuba}
\affiliation{Hiroshima Institute of Technology, Hiroshima}
\affiliation{University of Illinois at Urbana-Champaign, Urbana, Illinois 61801}
\affiliation{India Institute of Technology Guwahati, Guwahati}
\affiliation{Institute of High Energy Physics, Chinese Academy of Sciences, Beijing}
\affiliation{Institute of High Energy Physics, Vienna}
\affiliation{Institute of High Energy Physics, Protvino}
\affiliation{Institute of Mathematical Sciences, Chennai}
\affiliation{INFN - Sezione di Torino, Torino}
\affiliation{Institute for Theoretical and Experimental Physics, Moscow}
\affiliation{J. Stefan Institute, Ljubljana}
\affiliation{Kanagawa University, Yokohama}
\affiliation{Institut f\"ur Experimentelle Kernphysik, Universit\"at Karlsruhe, Karlsruhe}
\affiliation{Korea University, Seoul}
\affiliation{Kyoto University, Kyoto}
\affiliation{Kyungpook National University, Taegu}
\affiliation{\'Ecole Polytechnique F\'ed\'erale de Lausanne (EPFL), Lausanne}
\affiliation{Faculty of Mathematics and Physics, University of Ljubljana, Ljubljana}
\affiliation{University of Maribor, Maribor}
\affiliation{Max-Planck-Institut f\"ur Physik, M\"unchen}
\affiliation{University of Melbourne, School of Physics, Victoria 3010}
\affiliation{Nagoya University, Nagoya}
\affiliation{Nara University of Education, Nara}
\affiliation{Nara Women's University, Nara}
\affiliation{National Central University, Chung-li}
\affiliation{National United University, Miao Li}
\affiliation{Department of Physics, National Taiwan University, Taipei}
\affiliation{H. Niewodniczanski Institute of Nuclear Physics, Krakow}
\affiliation{Nippon Dental University, Niigata}
\affiliation{Niigata University, Niigata}
\affiliation{University of Nova Gorica, Nova Gorica}
\affiliation{Novosibirsk State University, Novosibirsk}
\affiliation{Osaka City University, Osaka}
\affiliation{Osaka University, Osaka}
\affiliation{Panjab University, Chandigarh}
\affiliation{Peking University, Beijing}
\affiliation{Princeton University, Princeton, New Jersey 08544}
\affiliation{RIKEN BNL Research Center, Upton, New York 11973}
\affiliation{Saga University, Saga}
\affiliation{University of Science and Technology of China, Hefei}
\affiliation{Seoul National University, Seoul}
\affiliation{Shinshu University, Nagano}
\affiliation{Sungkyunkwan University, Suwon}
\affiliation{School of Physics, University of Sydney, NSW 2006}
\affiliation{Tata Institute of Fundamental Research, Mumbai}
\affiliation{Excellence Cluster Universe, Technische Universit\"at M\"unchen, Garching}
\affiliation{Toho University, Funabashi}
\affiliation{Tohoku Gakuin University, Tagajo}
\affiliation{Tohoku University, Sendai}
\affiliation{Department of Physics, University of Tokyo, Tokyo}
\affiliation{Tokyo Institute of Technology, Tokyo}
\affiliation{Tokyo Metropolitan University, Tokyo}
\affiliation{Tokyo University of Agriculture and Technology, Tokyo}
\affiliation{Toyama National College of Maritime Technology, Toyama}
\affiliation{IPNAS, Virginia Polytechnic Institute and State University, Blacksburg, Virginia 24061}
\affiliation{Yonsei University, Seoul}
  \author{I.~Adachi}\affiliation{High Energy Accelerator Research Organization (KEK), Tsukuba} 
  \author{H.~Aihara}\affiliation{Department of Physics, University of Tokyo, Tokyo} 
  \author{K.~Arinstein}\affiliation{Budker Institute of Nuclear Physics, Novosibirsk}\affiliation{Novosibirsk State University, Novosibirsk} 
  \author{T.~Aso}\affiliation{Toyama National College of Maritime Technology, Toyama} 
  \author{V.~Aulchenko}\affiliation{Budker Institute of Nuclear Physics, Novosibirsk}\affiliation{Novosibirsk State University, Novosibirsk} 
  \author{T.~Aushev}\affiliation{\'Ecole Polytechnique F\'ed\'erale de Lausanne (EPFL), Lausanne}\affiliation{Institute for Theoretical and Experimental Physics, Moscow} 
  \author{T.~Aziz}\affiliation{Tata Institute of Fundamental Research, Mumbai} 
  \author{S.~Bahinipati}\affiliation{University of Cincinnati, Cincinnati, Ohio 45221} 
  \author{A.~M.~Bakich}\affiliation{School of Physics, University of Sydney, NSW 2006} 
  \author{V.~Balagura}\affiliation{Institute for Theoretical and Experimental Physics, Moscow} 
% ...
\collaboration{The Belle Collaboration}

\collaboration{The Belle Collaboration}

\begin{abstract}
  We present a measurement of the Cabibbo-Kobayashi-Maskawa matrix
element~$|V_{cb}|$ times the form factor normalization
$\mathcal{F}(1)$ using the decay $B^+\to\bar
D^{*0}\ell^+\nu_\ell$. This measurement is performed together with a
determination of the form factor parameters $\rho^2$, $R_1(1)$ and
$R_2(1)$ which fully characterize this decay in the framework of Heavy
Quark Effective Theory. This analysis is based on a data sample
equivalent to 140 fb$^{-1}$ of Belle data collected near the
$\Upsilon(4S)$~resonance.

The preliminary results, based on about 27,000
reconstructed $B^+\to\bar D^{*0}\ell^+\nu_\ell$ decays, are
$\rho^2=1.376\pm 0.074\pm 0.056$, $R_1(1)=1.620\pm 0.091\pm 0.092$,
$R_2(1)=0.805\pm 0.064\pm 0.036$ and $\mathcal{F}(1)|V_{cb}|=35.0\pm
0.4\pm 2.2$. We find the $B^+\to\bar D^{*0}\ell^+\nu_\ell$ branching
fraction to be $(4.84\pm 0.04\pm 0.56)\%$. For all numbers quoted
here, the first error is the statistical and the second is the
systematic uncertainty.

A direct, model-independent determination of the form factor shapes
has also been carried out and shows good agreement with the HQET based
form factor parametrization by Caprini {\it et al.}.

\end{abstract}


\maketitle


{\renewcommand{\thefootnote}{\fnsymbol{footnote}}}
\setcounter{footnote}{0}

\section{Introduction}

The magnitude of the Cabibbo-Kobayashi-Maskawa (CKM) matrix element
$|V_{cb}|$ can be determined from exclusive decays $B\to\bar
D^*\ell\nu$~\cite{Kobayashi:1973fv}. In the framework of Heavy Quark
Effective Theory (HQET), this decay is described by three form factor
parameters, $\rho^2$, $R_1(1)$ and
$R_2(1)$~\cite{Neubert:1993mb,Caprini:1997mu} along with a form factor
normalization $\mathcal{F}(1)$ derived from lattice
QCD~\cite{Bernard:2008dn}. Many measurements of the $B^0$~mode are
available which are unfortunately barely
consistent~\cite{Barberio:2008fa}, suggesting a hidden systematic
uncertainty.

To shed light on this inconsistency, we now investigate the related
decay $B^+\to\bar D^{*0}\ell^+\nu_\ell$ which carries
different experimental systematics. This is interesting
also because only few measurements of the $B^+$~mode are available in
the literature~\cite{Albrecht:1991iz,Adam:2002uw,Aubert:2007qs}.

This paper is organized as follows: after introducing the theoretical
framework for the study of this decay, the experimental procedure is
presented in detail. This is followed by a discussion of our results
and the systematic uncertainties. Finally, a form factor
parameterization independent measurement of the form factor shapes is
described and results are given.

\section{Theoretical framework}

\subsection{Kinematic variables} \label{sec:2a}

\begin{figure}
  \begin{center}
    \includegraphics[width=5cm]{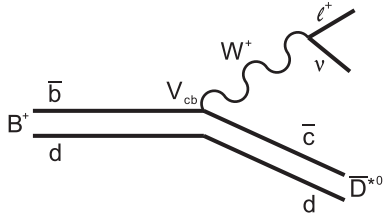}
  \end{center}
  \caption{Quark-level Feynman diagram for the decay $B^+\to\bar
  D^{*0}\ell^+\nu_\ell$.} \label{fig:1}
\end{figure}

The decay $B^+\to\bar D^{*0}\ell^+\nu_\ell$~\cite{ref:0} proceeds
chiefly through the tree-level transition shown in
Fig.~\ref{fig:1}.
Its kinematics can be fully characterized by four variables:

The first one is $w$, defined by
\begin{equation}
  w=\frac{P_B\cdot P_{D^{*0}}}{m_B
    m_{D^{*0}}}=\frac{m_B^2+m_{D^{*0}}^2-q^2}{2m_Bm_{D^{*0}}}~,
\end{equation}
where $m_B$ and $m_{D^{*0}}$ are the masses of the $B^+$ and the $D^{*0}$
mesons (5.2792 GeV/$c^2$ and 2.007 GeV/$c^2$, respectively~\cite{Amsler:2008zzb}),
$P_B$ and $P_{D^{*0}}$ are their four-momenta, and
$q^2=(P_\ell+P_\nu)^2$. In the $B$~rest frame, approximately equal to
the $\Upsilon(4S)$~center-of-mass (c.m.) frame, the expression for $w$
reduces to the Lorentz boost
$\gamma_{D^{*0}}=E_{D^{*0}}/m_{D^{*0}}$. The ranges of $w$ and $q^2$
are restricted by the kinematics of the decay, with $q^2 = 0$
corresponding to
\begin{equation}
  w_\mathrm{max}=\frac{m_B^2 + m_{D^{*0}}^2}{2m_B m_{D^{*0}}} \approx
  1.505~,
\end{equation}
and $w_\mathrm{min}=1$ to
\begin{equation}
  q_\mathrm{max}^2=(m_B-m_{D^{*0}})^2 \approx 10.71~\mathrm{GeV}^2/c^4~.
\end{equation}
The point~$w=1$ is also refered to as zero recoil.

\begin{figure}
  \begin{center}
    \includegraphics[width=0.8\columnwidth]{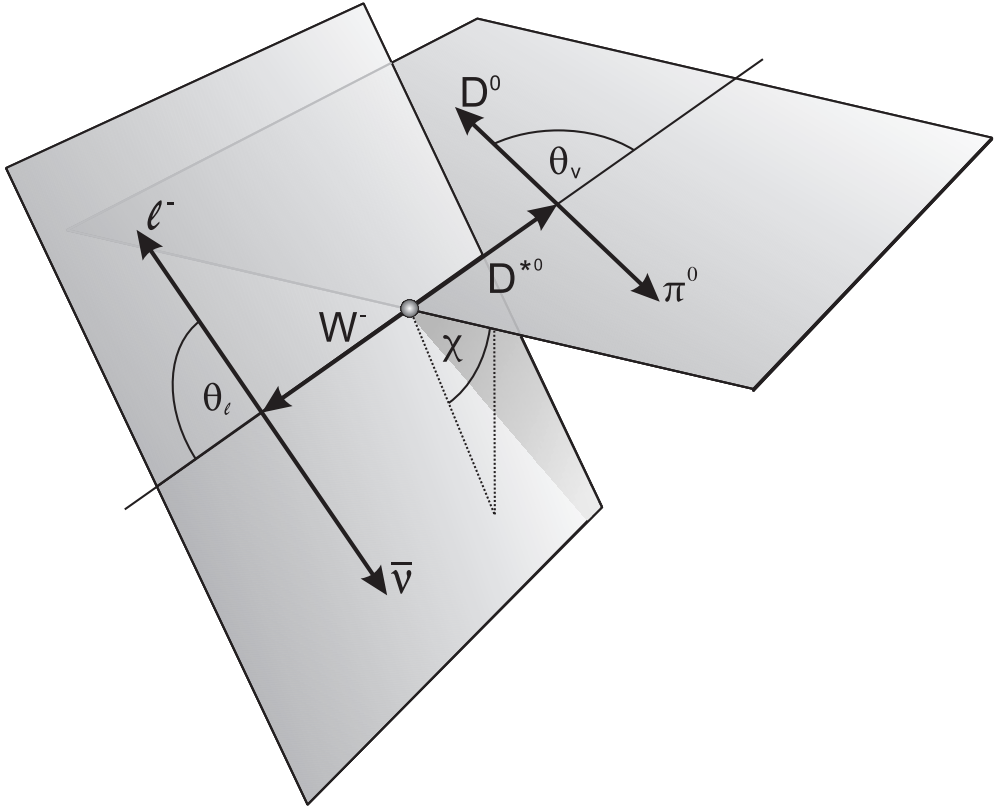}
  \end{center}
  \caption{Definition of the angles $\theta_\ell$, $\theta_V$ and
    $\chi$ for the decay $B^+\to\bar D^{*0}\ell^+\nu_\ell$, $D^{*0}\to
    D^0\pi_s^0$.} \label{fig:2}
\end{figure}

The remaining three variables are the angles shown in Fig.~\ref{fig:2}:
\begin{itemize} 
\item{$\theta_\ell$, the angle between the direction of the lepton
  in the virtual $W$~rest frame and the direction of the $W$ in the
  $B$~rest frame;}
\item{$\theta_V$, the angle between the direction of the $D$~meson in the
  $D^{*0}$ rest frame and the direction of the $D^{*0}$~meson in the $B$
  rest frame;}
\item{$\chi$, the angle between the plane formed by the $D^{*0}$ and
  the plane formed by the $W$~decay.}
\end{itemize} 

\subsection{Four-dimensional decay distribution}

The Lorentz structure of the $B^+\to\bar D^{*0}\ell^+\nu_\ell$~decay
amplitude can be expressed in terms of three helicity amplitudes
($H_{+}$, $H_{-}$, and $H_{0}$), which correspond to the three
polarization states of the $D^{*0}$, two transverse and one
longitudinal. For low-mass leptons (electrons and muons), these
amplitudes are expressed in terms of the three functions $h_{A_1}(w)$,
$R_1(w)$, and $R_2(w)$~\cite{Neubert:1993mb}
\begin{equation}
  H_i(w)=m_B\frac{R^*(1-r^2)(w+1)}{2\sqrt{1-2wr+r^2}}h_{A_1}(w)\tilde{H}_i(w)~,
\end{equation}
where
\begin{eqnarray}
  \tilde{H}_{\mp} & = &
  \frac{\sqrt{1-2wr+r^2}\left(1\pm\sqrt{\frac{w-1}{w+1}}
    R_1(w)\right)}{1-r}~, \\
  \tilde{H}_0 & = & 1+\frac{(w-1)(1-R_2(w))}{1-r}~,
\end{eqnarray}
with $R^*=(2\sqrt{m_B m_{D^{*0}}})/(m_B+m_{D^{*0}})$ and
$r=m_{D^{*0}}/m_B$. The functions $R_1(w)$ and $R_2(w)$ are defined in
terms of the axial and vector form factors as,
\begin{equation}
  A_2(w)=\frac{R_2(w)}{R^{*2}}\frac{2}{w+1}A_1(w)~,
\end{equation}
\begin{equation}
  V(w)=\frac{R_1(w)}{R^{*2}}\frac{2}{w+1}A_1(w)~.
\end{equation}
By convention, the function $h_{A_1}(w)$ is defined as
\begin{equation}
  h_{A_1}(w)=\frac{1}{R^*}\frac {2}{w+1} A_1(w)~.
\end{equation}
For $w\to 1$, the axial form factor $A_1(w)$ dominates, and in the
limit of infinite $b$- and $c$-quark masses, a single form factor
describes the decay, the so-called Isgur-Wise
function~\cite{Isgur:1989vq,Isgur:1989ed}.

The fully differential decay rate in terms of the three helicity
amplitudes is
\begin{equation}
  \begin{split}
    & \frac{\mathrm{d}^4\Gamma(B^+\to\bar
    D^{*0}\ell^+\nu_\ell)}{\mathrm{d}w\mathrm{d}(\cos\theta_\ell)\mathrm{d}(\cos\theta_V)\mathrm{d}\chi}=\frac{6m_Bm_{D^{*0}}^2}{8(4\pi)^4}\sqrt{w^2-1}(1-2wr+r^2)G_F^2|V_{cb}|^2\\
    & \times\big\{(1-\cos\theta_\ell)^2\sin^2\theta_VH^2_+(w)+(1+\cos\theta_\ell)^2\sin^2\theta_VH^2_-(w)\biggr. \\
    & +4\sin^2\theta_\ell\cos^2\theta_VH^2_0(w)-2\sin^2\theta_\ell\sin^2\theta_V\cos 2\chi H_+(w)H_-(w) \\
    & -4\sin\theta_\ell(1-\cos\theta_\ell)\sin\theta_V\cos\theta_V\cos\chi H_+(w)H_0(w) \\
    &
      +\biggl. 4\sin\theta_\ell(1+\cos\theta_\ell)\sin\theta_V\cos\theta_V\cos\chi H_-(w)H_0(w)\big\}~, \label{eq:2_1}
  \end{split}
\end{equation}
with $G_F=(1.16637\pm 0.00001)\times 10^{-5}$~GeV$^{-2}$~\cite{Amsler:2008zzb}. By
integrating this decay rate over all but one of the four variables,
$w$, $\cos\theta_\ell$, $\cos\theta_V$, or $\chi$, we obtain the four
one-dimensional decay distributions from which we will extract the
form factors. The differential decay rate as a function of $w$ is
\begin{equation}
  \frac{\mathrm{d}\Gamma}{\mathrm{d}w}=\frac{G^2_F}{48\pi^3}m^3_{D^{*0}}\big(m_B-m_{D^{*0}}\big)^2\mathcal{G}(w)\mathcal{F}^2(w)|V_{cb}|^2~, \label{eq:2_2}
\end{equation}
where  
\begin{eqnarray*}
  \mathcal{F}^2(w)\mathcal{G}(w)=h_{A_1}^2(w)\sqrt{w-1}(w+1)^2\left\{2\left[\frac{1-2wr+r^2}{(1-r)^2}\right]\right.
  \\
  \left. \times\left[1+R_1(w)^2\frac{w-1}{w+1}\right]+\left[1+(1-R_2(w))\frac{w-1}{1-r}\right]^2\right\}~,
\end{eqnarray*}
and $\mathcal{G}(w)$ is a known phase space factor,
\begin{equation*}
  \mathcal{G}(w)=\sqrt{w^2-1}(w+1)^2\left[1+4\frac{w}{w+1}\frac{1-2wr+r^2}{(1-r)^2}\right].
\end{equation*}

In the infinite quark-mass limit, the heavy quark symmetry (HQS)
predicts $\mathcal{F}(1)=1$. Corrections to this limit have been
calculated in lattice QCD. The most recent result obtained in
unquenched lattice QCD reads $\mathcal{F}(1)=0.921\pm 0.013\pm
0.020$~\cite{Bernard:2008dn}.

\subsection{Form factor parameterization} \label{sub:parametrization}

The heavy quark effective theory (HQET) allows a parameterization of 
these form-factors to be obtained. Perfect heavy quark symmetry
implies that $R_1(w)=R_2(w)=1$, {\it i.e.}, the form factors $A_2$ and
$V$ are identical for all values of $w$ and differ from $A_1$ only by
a simple kinematic factor. Corrections to this approximation have been
calculated in powers of $\Lambda_\mathrm{QCD}/m_b$ and the strong
coupling constant $\alpha_s$. Various parameterizations in powers of
$(w-1)$ have been proposed. Among the different predictions relating
the coefficients of the higher order terms to the linear term, we
adopt the following expressions derived by Caprini, Lellouch and
Neubert~\cite{Caprini:1997mu},
\begin{eqnarray}
  h_{A_1}(w) & = &
  h_{A_1}(1)\big[1-8\rho^2z+(53\rho^2-15)z^2-(231\rho^2-91)z^3\big]~,
  \label{eq:2_3} \\
  R_1(w) & = & R_1(1)-0.12(w-1)+0.05(w-1)^2~, \label{eq:2_4} \\ 
  R_2(w) & = & R_2(1)+0.11(w-1)-0.06(w-1)^{2}~, \label{eq:2_5}  
\end{eqnarray}
where $z=(\sqrt{w+1}-\sqrt{2})/(\sqrt{w+1}+\sqrt{2})$. The three
parameters $\rho^{2}$, $R_1(1)$, and $R_2(1)$, cannot be calculated;
they must be extracted from data.

\section{Experimental procedure}

\subsection{Data sample and event selection}

The data used in this analysis were taken with the Belle
detector~\cite{unknown:2000cg} at the KEKB asymmetric energy
$e^+e^-$~collider~\cite{Kurokawa:2001nw}. Belle is a large-solid-angle
magnetic spectrometer that consists of a silicon vertex
detector (SVD), a 50-layer central drift chamber (CDC), an array of
aerogel threshold Cherenkov counters (ACC), a barrel-like
arrangement of time-of-flight scintillation counters (TOF), and an
electromagnetic calorimeter (ECL) comprised of CsI(Tl) crystals
located inside a super-conducting solenoid coil that provides a 1.5~T
magnetic field. An iron flux-return located outside of the coil is
instrumented to detect $K_L^0$ mesons and to identify muons (KLM).
In the original setup the SVD was composed of three layers of double sided
silicon strip modules, in 2003 it was upgraded and a fourth layer
was added.

The data sample consists of 140~fb$^{-1}$ taken at the
$\Upsilon(4S)$~resonance, or $152\times 10^6$ $B\bar B$~events. The 
entire data sample has been recorded using the three-layer SVD setup. 
Another 15~fb$^{-1}$ taken at 60~MeV below the resonance are
used to estimate the non-$B\bar B$ (continuum) background. The
off-resonance data is scaled by the integrated on- to off-resonance
luminosity ratio corrected for the $1/s$~dependence of the $q\bar
q$~cross-section. This sample is identical to the one used for
the $B^0\to D^{*-}\ell^+\nu_\ell$~analysis~\cite{:2008nd}.

Monte Carlo (MC) samples equivalent to about three times the
integrated luminosity are used in this analysis. MC simulated
events are generated with the evtgen program~\cite{Lange:2001uf} and
full detector simulation based on GEANT~\cite{Brun:1987ma} is
applied. QED bremsstrahlung in $B\to X\ell\nu$~decays is added using
the PHOTOS package~\cite{Barberio:1993qi}.

Hadronic events are selected based on the charged track multiplicity
and the visible energy in the calorimeter. The selection is described
in detail elsewhere~\cite{Abe:2001hj}. We also apply a moderate cut on
the ratio of the second to the zeroth Fox-Wolfram
moment~\cite{Fox:1978vu}, $R_2<0.4$, to reject continuum events.

\subsection{Event reconstruction}

Charged tracks are required to originate from the interaction point by
applying the following selections on the impact parameters in $R\phi$
and $z$, $dr<2$~cm and $|dz|<4$~cm, respectively. Additionally, we
demand at least one associated hit in the SVD detector. For pion and
kaon candidates, the Cherenkov light yield from ACC, the
time-of-flight information from TOF and $dE/dx$ from CDC are required
to be consistent with the respective mass hypothesis.

Neutral $D$~meson candidates are searched for in the decays channels
$D^0\to K^-\pi^+$ and $D^0\to K^-\pi^+\pi^-\pi^+$. We fit the charged
tracks to a common vertex and reject the $D^0$~candidate if the
$\chi^2$-probability is below $10^{-3}$. The momenta of the charged
tracks are re-evaluated at the vertex and the $D^0$ 4-momentum is
calculated as their sum. The reconstructed $D^0$~mass is required to
lie within $\pm 3$ standard deviations from the nominal mass~\cite{Amsler:2008zzb}, where
one sigma (measured from real data) is about 4.5~MeV/$c^2$ (4~MeV/$c^2$)
for the one (three) pion mode.

The $D^0$~candidate is combined with a slow neutral pion~$\pi^0_s$ to form a
$D^{*0}$~candidate. The $\pi^0_s$ is obtained by combining two photons
with $E_\gamma>100$~MeV for $\theta<32^\circ$, $E_\gamma>150$~MeV for
$\theta>130^\circ$, and $E_\gamma>50$~MeV for photons detected in the
barrel region. The invariant $\gamma\gamma$ mass has to lie within 3
standard deviations of the nominal $\pi^0$ mass, with $\sigma\approx
5$~MeV/$c^2$. Additionally, a mass constraint fit is performed and the fit
probability has to exceed $10^{-2}$. Continuum suppression is achieved
by requiring a $D^{*0}$~momentum less than 2.45~GeV/$c$ in the c.m.\
frame.

Finally, the $D^{*0}$~candidate is combined with a lepton (electron or
muon), appropriately charged with respect to the kaon. Electron
candidates are identified using the ratio of the energy detected in
the ECL to the track momentum, the ECL shower shape, position matching
between track and ECL cluster, the energy loss in the CDC and the
response of the ACC~counters. Muons are identified based on their
penetration range and transverse scattering in the KLM~detector. In
the momentum region relevant to this analysis, charged leptons are
identified with an efficiency of about 90\% and the probability to
misidentify a pion as an electron (muon) is 0.25\%
(1.4\%)~\cite{Hanagaki:2001fz,Abashian:2002bd}. No SVD hit requirement
is made for lepton tracks. In the lab frame, the (transverse) momentum
of the lepton is required to exceed 0.80~GeV/$c$ (0.65~GeV/$c$) in
case of electrons and 0.85~GeV/$c$ (0.75~GeV/$c$) in case of muons. We
also apply an upper lepton momentum cut at 2.4~GeV/$c$ in the c.m.\ frame
to reject continuum.

In electron events, we attempt bremsstrahlung recovery by searching
for photons in a cone of 3$^\circ$ around the electron track. If such
a photon is found it is merged with the electron and the sum of the
momenta is assumed to be the lepton momentum.

\begin{figure}
  \begin{center}
    \includegraphics[width=0.4\columnwidth]{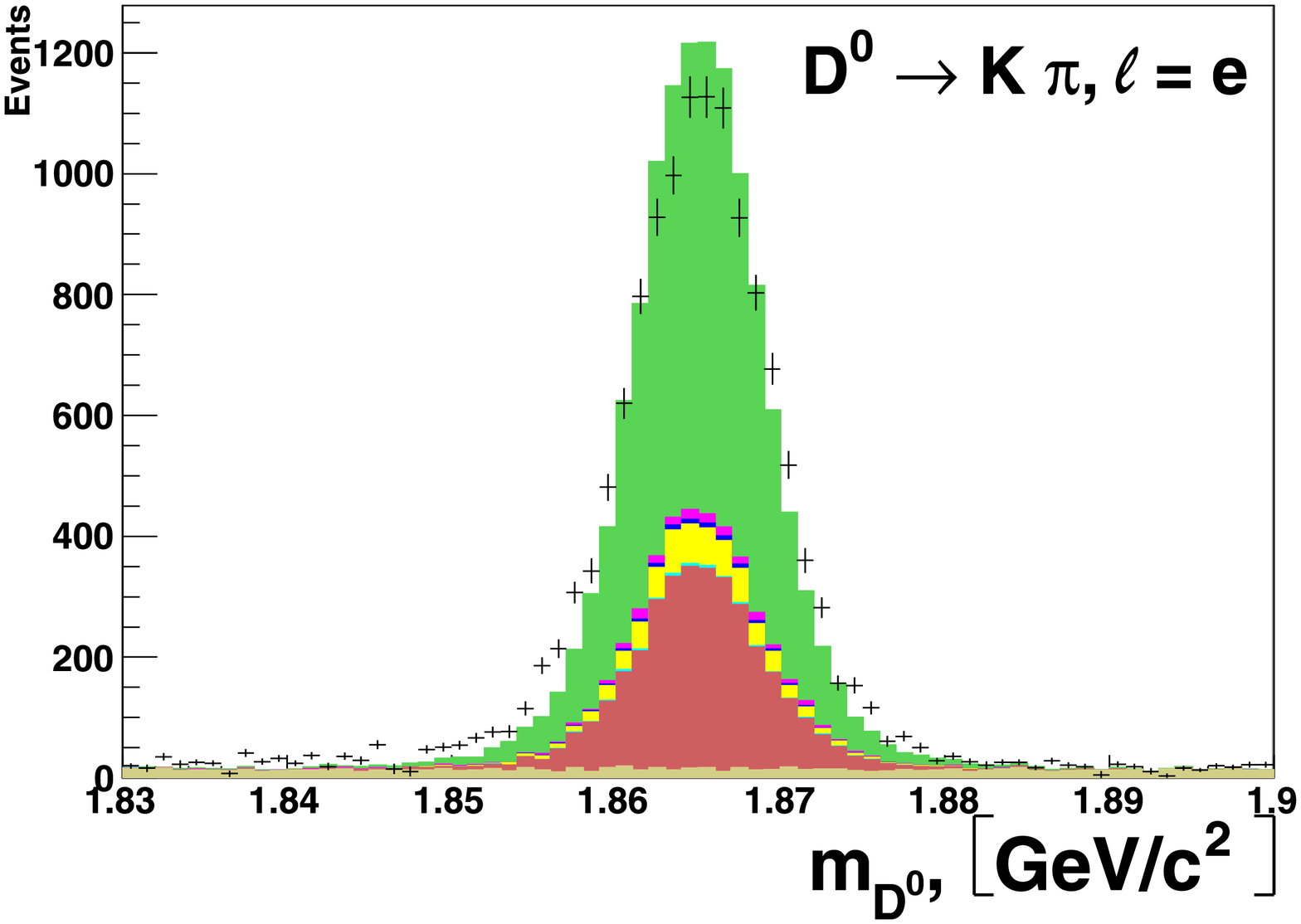}
    \includegraphics[width=0.4\columnwidth]{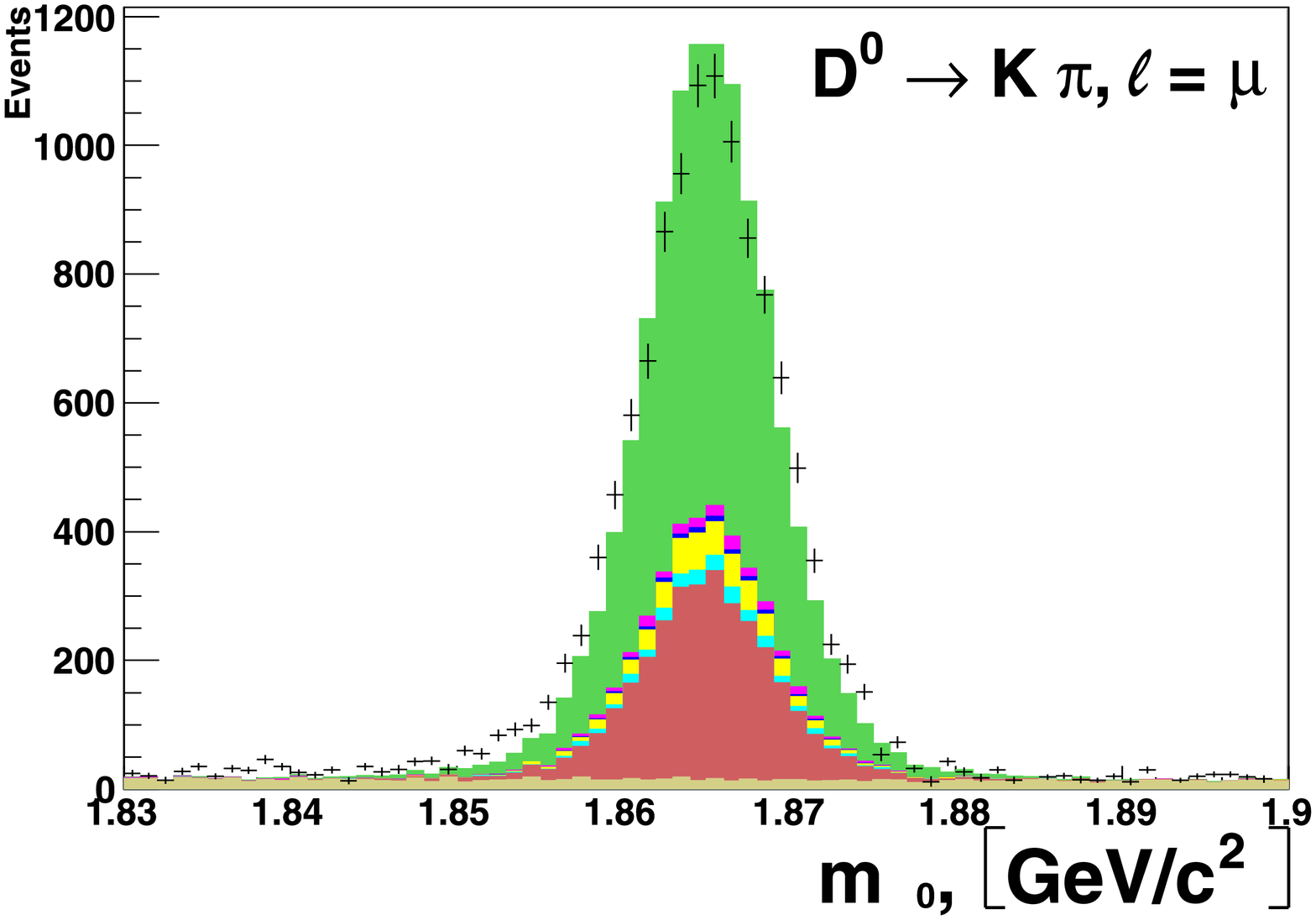}\\
    \vspace{0.01\columnwidth}
    \includegraphics[width=0.4\columnwidth]{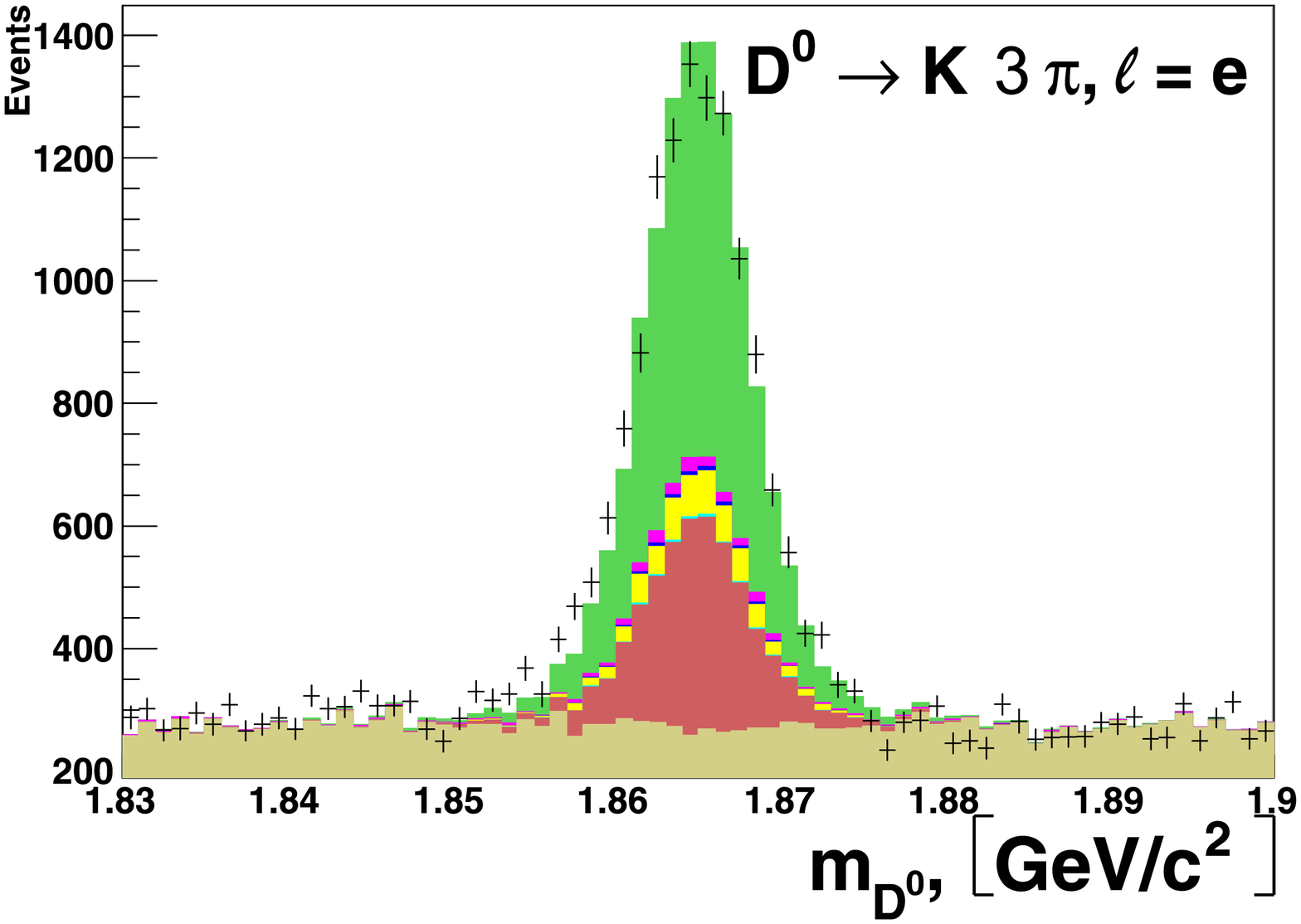}
    \includegraphics[width=0.4\columnwidth]{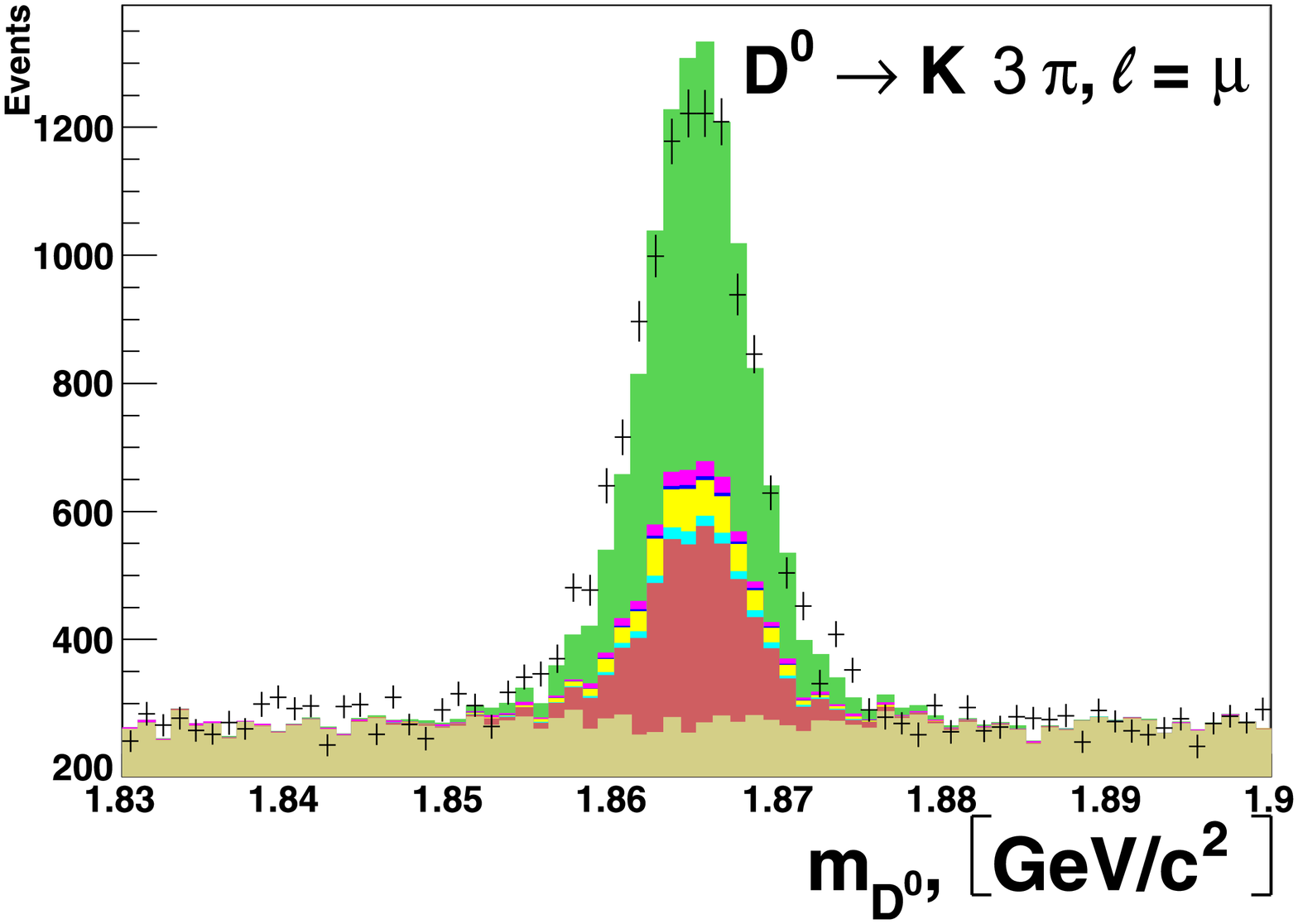}\\
    \includegraphics[width=0.4\columnwidth]{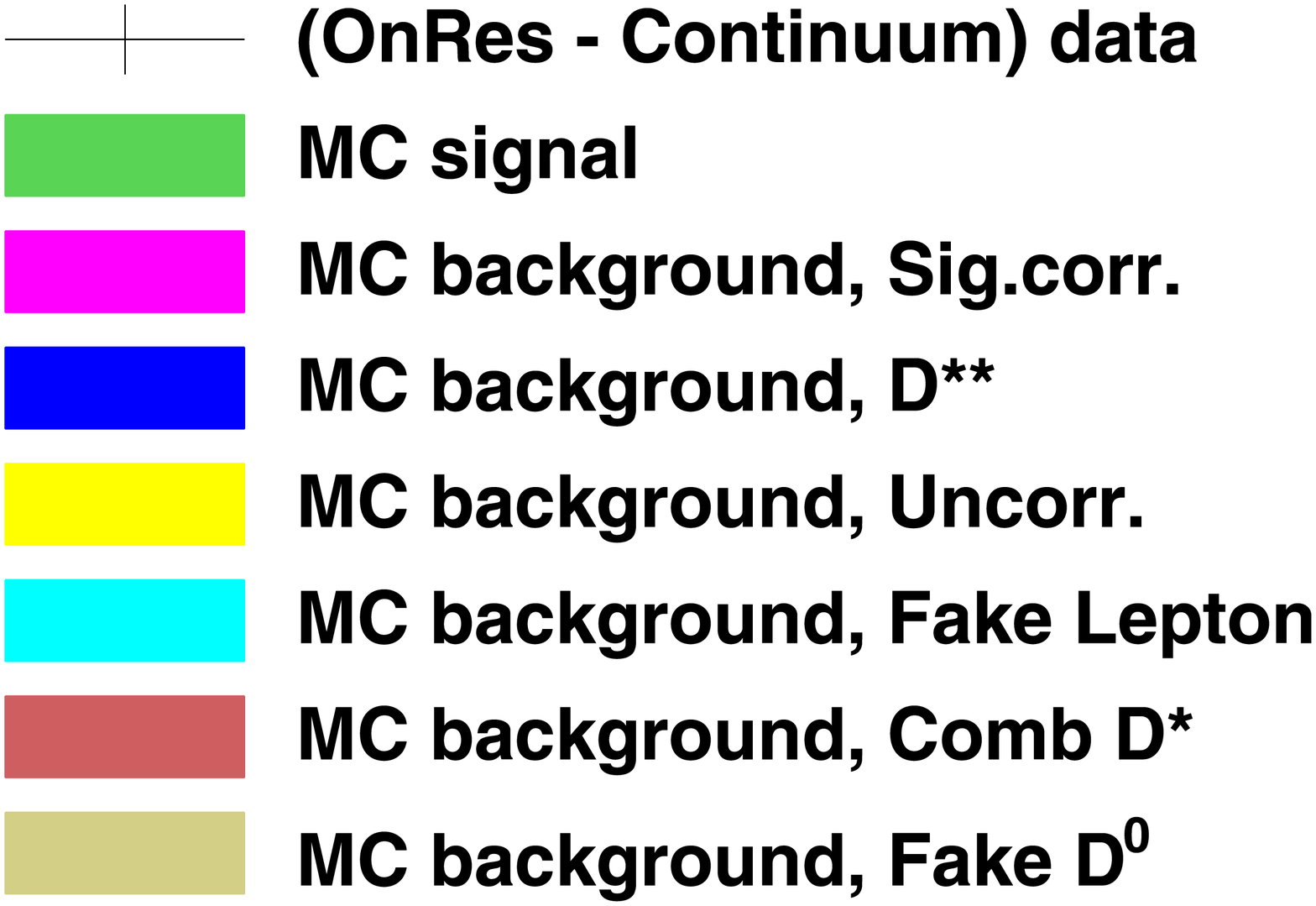}
  \end{center}
  \caption{Invariant $D^0$~candidate mass distributions in the
    different sub-samples. All analysis cuts (except on the plotted
    variable) are applied. The estimation of the background
    contributions is described in section~\ref{sec:3c}.}
    \label{fig:3}
\end{figure}

\begin{figure}
  \begin{center}
    \includegraphics[width=0.4\columnwidth]{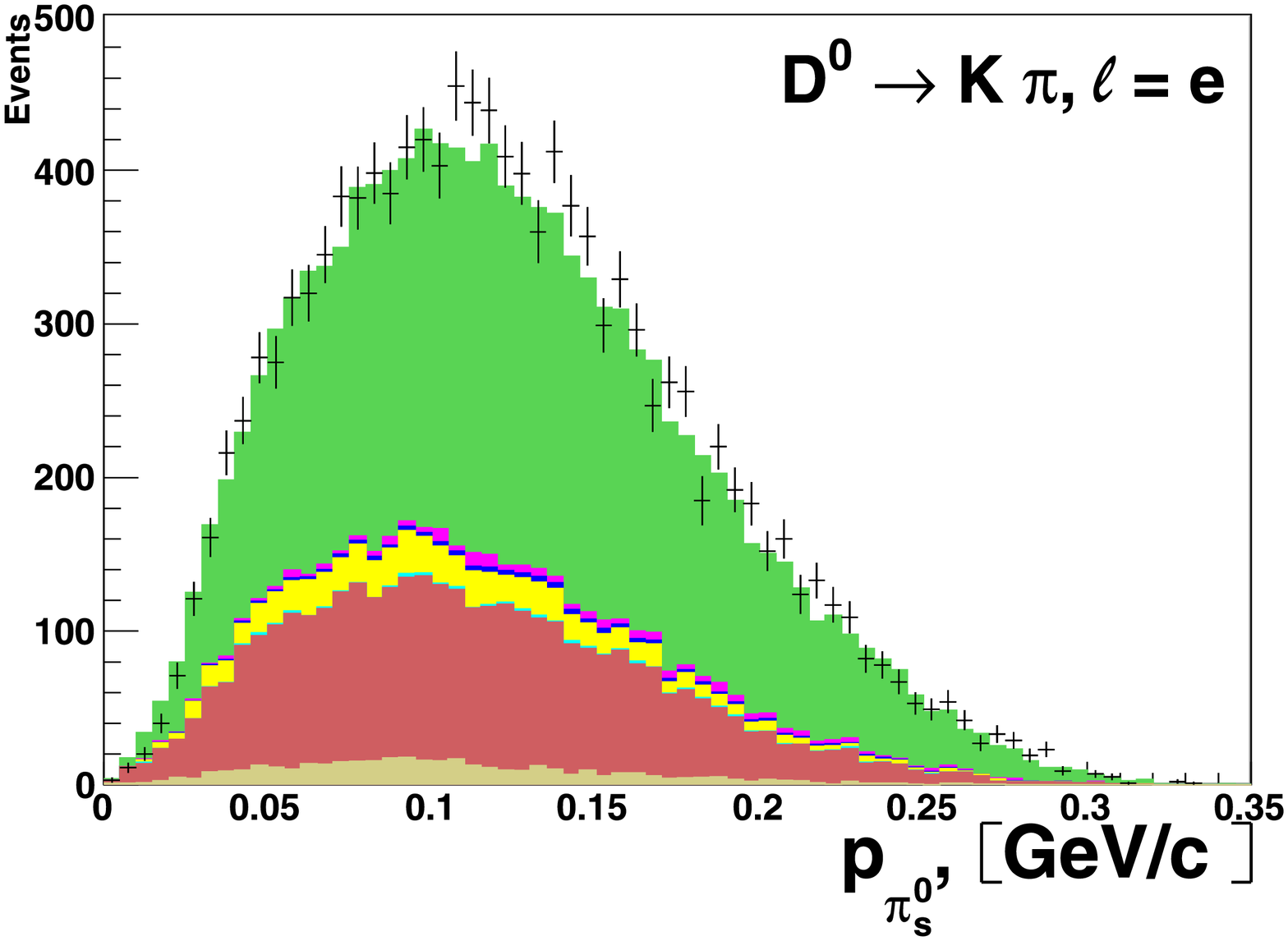}
    \includegraphics[width=0.4\columnwidth]{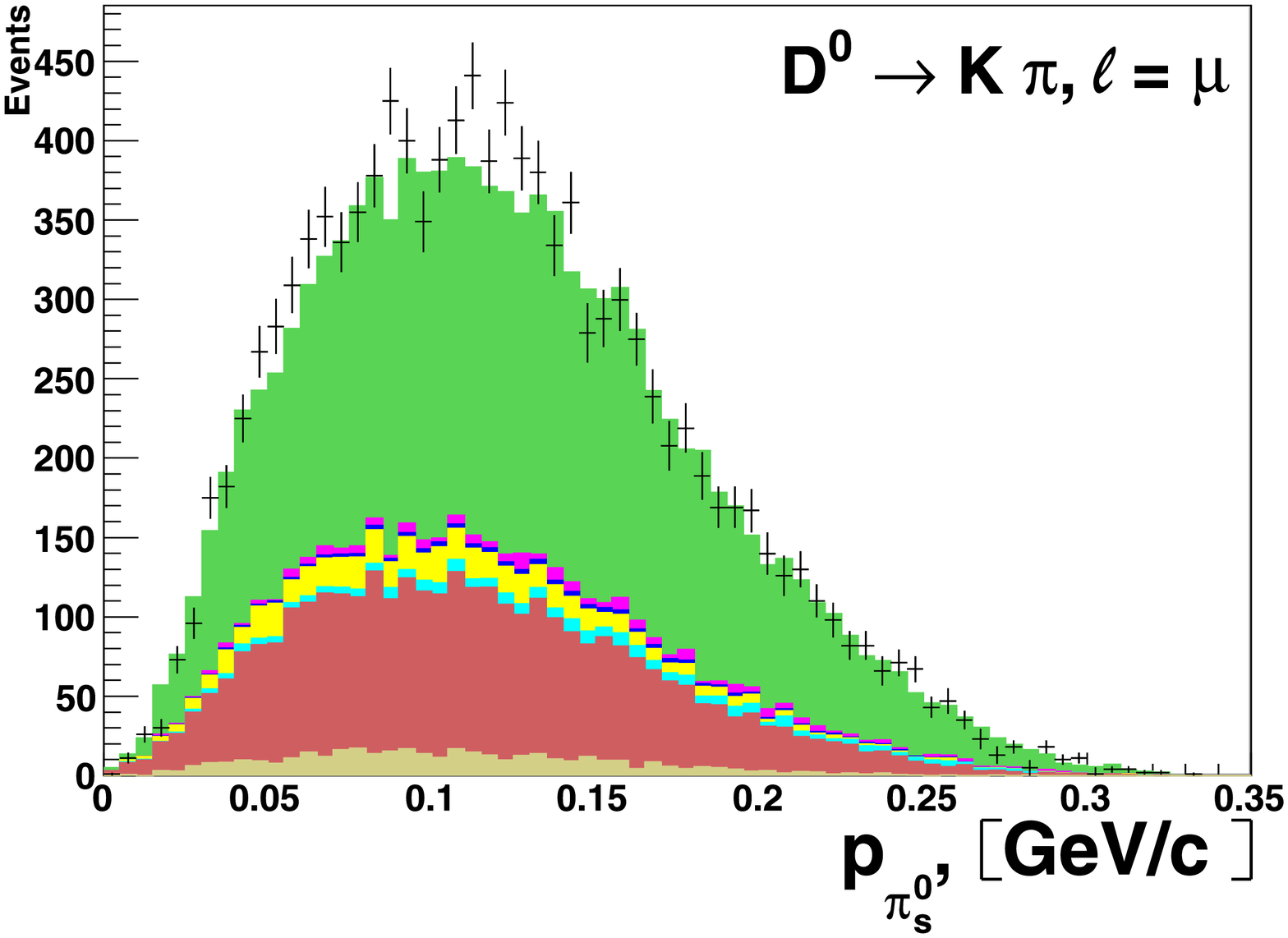}\\
    \vspace{0.01\columnwidth}
    \includegraphics[width=0.4\columnwidth]{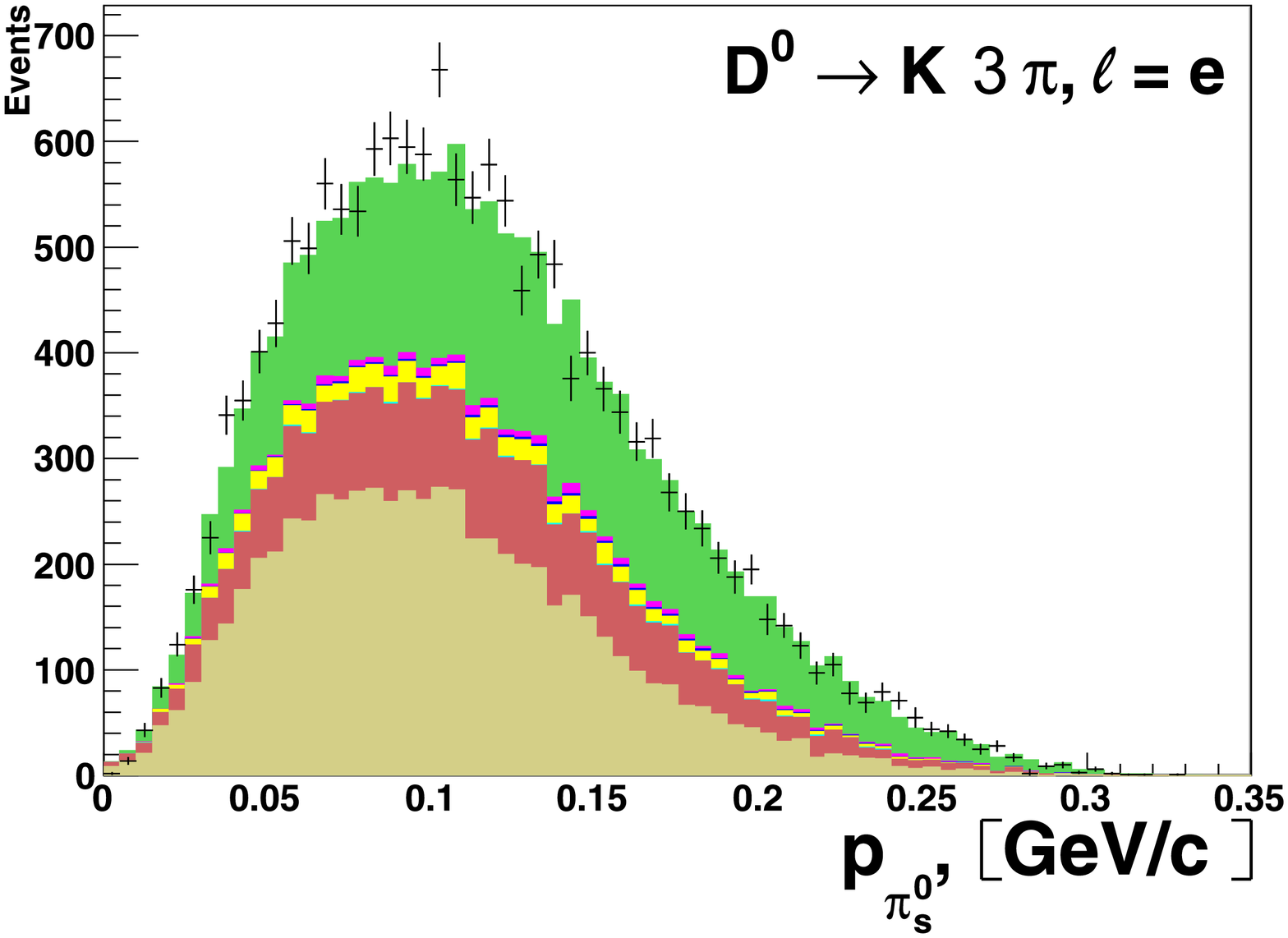}
    \includegraphics[width=0.4\columnwidth]{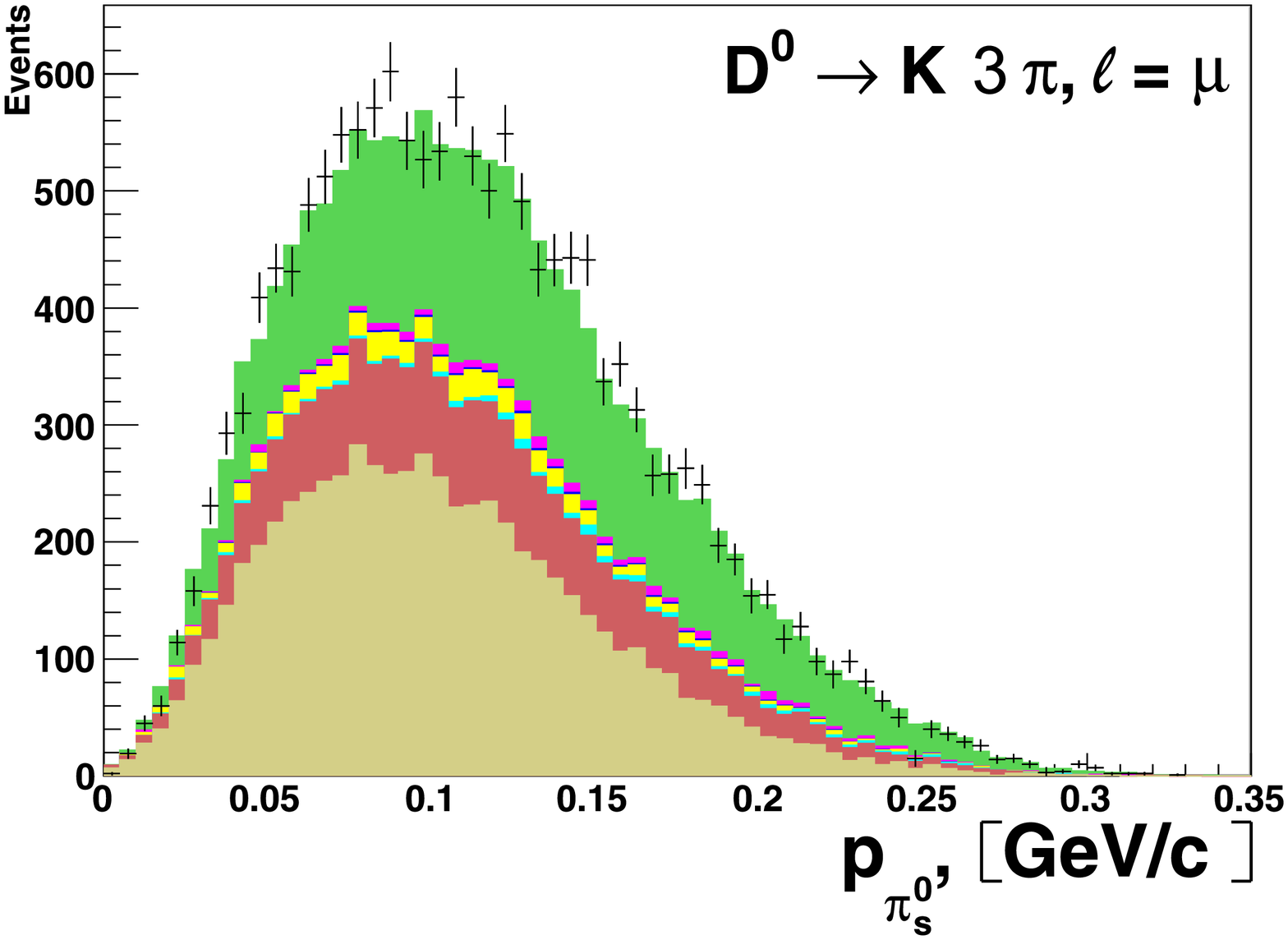}
  \end{center}
  \caption{Momentum distributions of the slow $\pi^0$ (lab frame) in
  the different sub-samples. All analysis cuts are applied. The color
  scheme is explained in Fig.~\ref{fig:3}. The estimation of the background
    contributions is described in section~\ref{sec:3c}.} \label{fig:4}
\end{figure}

Figs.~\ref{fig:3} and \ref{fig:4} show the invariant mass of the
$D^0$~candidates and the $\pi^0_s$ momentum distributions,
respectively.
 
\subsection{Background estimation} \label{sec:3c}

Because we do not reconstruct the other $B$~meson in the event, the
$B$ momentum is \emph{a priori} unknown. However, in the c.m.\ frame,
one can show that the $B$~direction lies on a cone around the
$(D^{*0}\ell)$-axis~\cite{ref:2},
\begin{equation}
  \cos\theta_{B,D^{*0}\ell}=\frac{2E^*_BE^*_{D^{*0}\ell}-m^2_B-m^2_{D^{*0}\ell}}{2|\vec
  p^*_B||\vec p^*_{D^{*0}\ell}|}~, \label{eq:3_1}
\end{equation}
where $E^*_B$ is half of the c.m.\ energy and $|\vec p^*_B|$ is
$\sqrt{E^{*2}_B-m^2_B}$. The quantities $E^*_{D^{*0}\ell}$, $\vec
p^*_{D^{*0}\ell}$ and $m_{D^{*0}\ell}$ are calculated from the
reconstructed $D^{*0}\ell$~system.

This cosine is also a powerful discriminator between signal and
background: signal events should strictly lie in the interval $(-1,1)$,
although -- due to finite detector resolution -- about 5\% of the signal is
reconstructed outside this interval. The background on the other hand
does not have this restriction.

To determine the amount of signal and background in the selected
events, we perform a two dimensional binned fit to
$\cos\theta_{B,D^{*0}\ell}$ and $\Delta m$, the invariant mass
difference between the $D^{*0}$ and the $D$~candidates. We consider
backgrounds from the following six sources:
\begin{itemize}
\item continuum: any candidate reconstructed in a non-$\Upsilon(4S)$
  event,

\item fake $D^0$: the $D^0$~candidate has been misreconstructed,

\item combinatoric $D^{*0}$: the $D^{*0}$~candidate is
  misreconstructed, however the $D^0$ candidate is identified
  correctly,

\item fake lepton: the lepton has been misidentified but the $D^{*0}$
  is reconstructed properly,

\item uncorrelated background: the $D^{*0}$ and the lepton stem from
  different $B$~mesons,

\item $D^{**}$: background from $B\to\bar D^{**}\ell^+\nu$~decays with
    $\bar D^{**}\to\bar D^{*0}\pi$ or $B\to\bar D^{*0}\pi\ell^+\nu$
    non-resonant, and

\item correlated background: background from other processes in
    which the $D^{*0}$ and the lepton stem from the same $B$~meson, {\it e.g},
    $B^+\to\bar D^{*0}\tau^+\nu$, $\tau^+\to\mu^+\nu\nu$.
\end{itemize}

\begin{figure}
  \begin{center}
    \includegraphics[width=0.45\columnwidth]{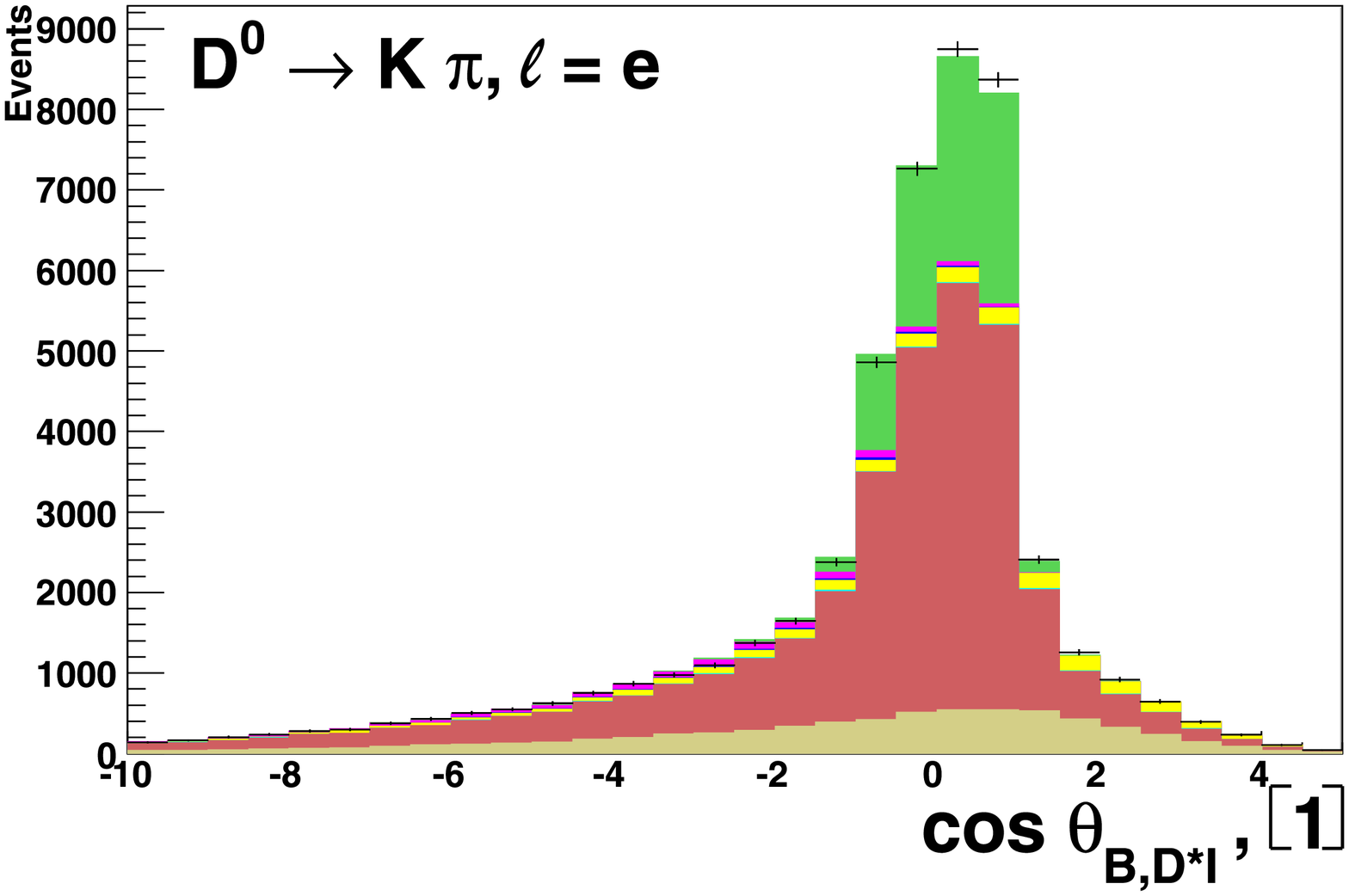}
    \includegraphics[width=0.45\columnwidth]{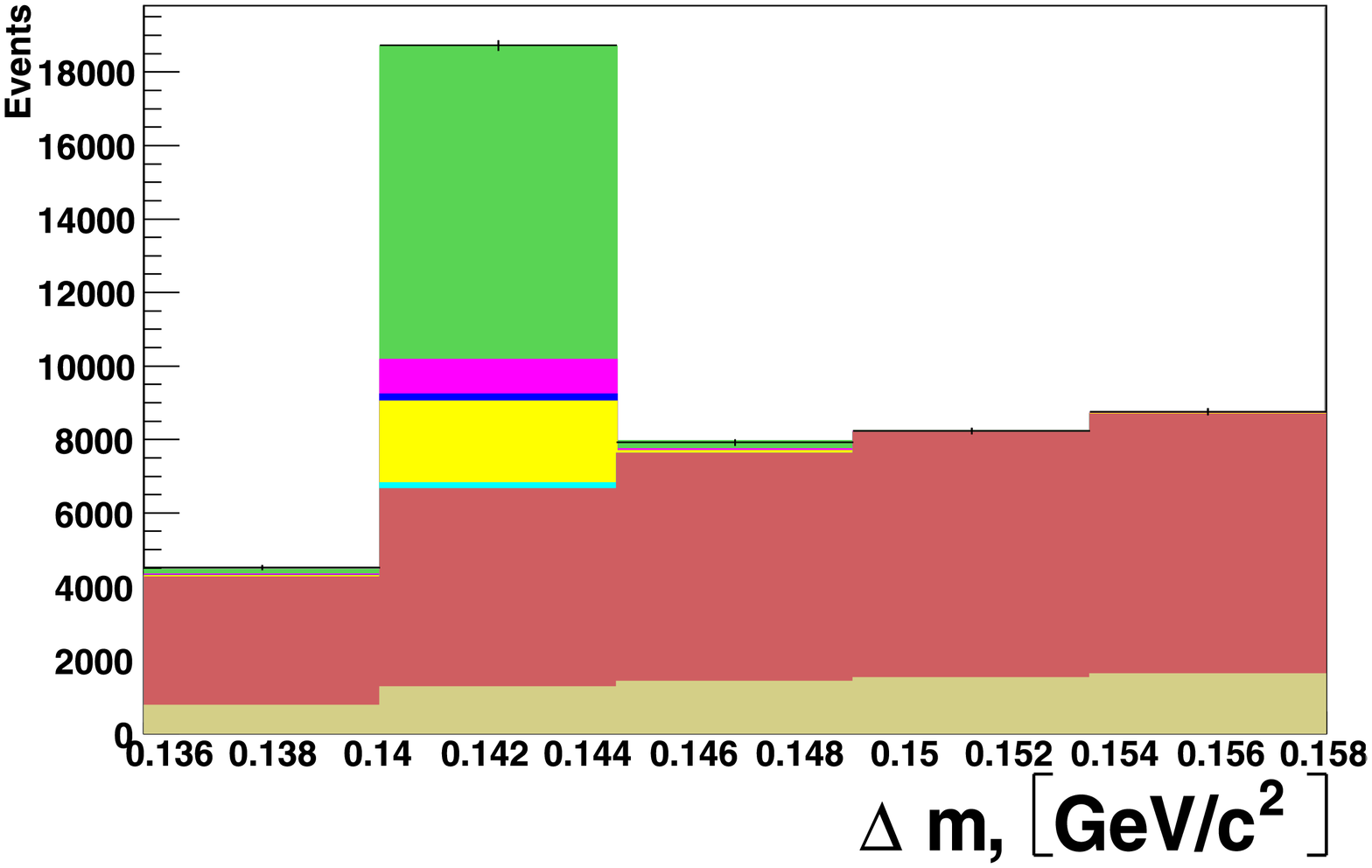}
  \end{center}
  \caption{Result of the fits to the $\cos\theta_{B,D^{*0}\ell}$ vs.\
    $\Delta m$~distributions in the $K\pi, e$ sub-sample. The
    projections in $\cos\theta_{B,D^{*0}\ell}$ 
    and $\Delta m$ are shown. The color
  	scheme is explained in Fig.~\ref{fig:3}.} \label{fig:5a}
\end{figure}

\begin{figure}
  \begin{center}
    \includegraphics[width=0.45\columnwidth]{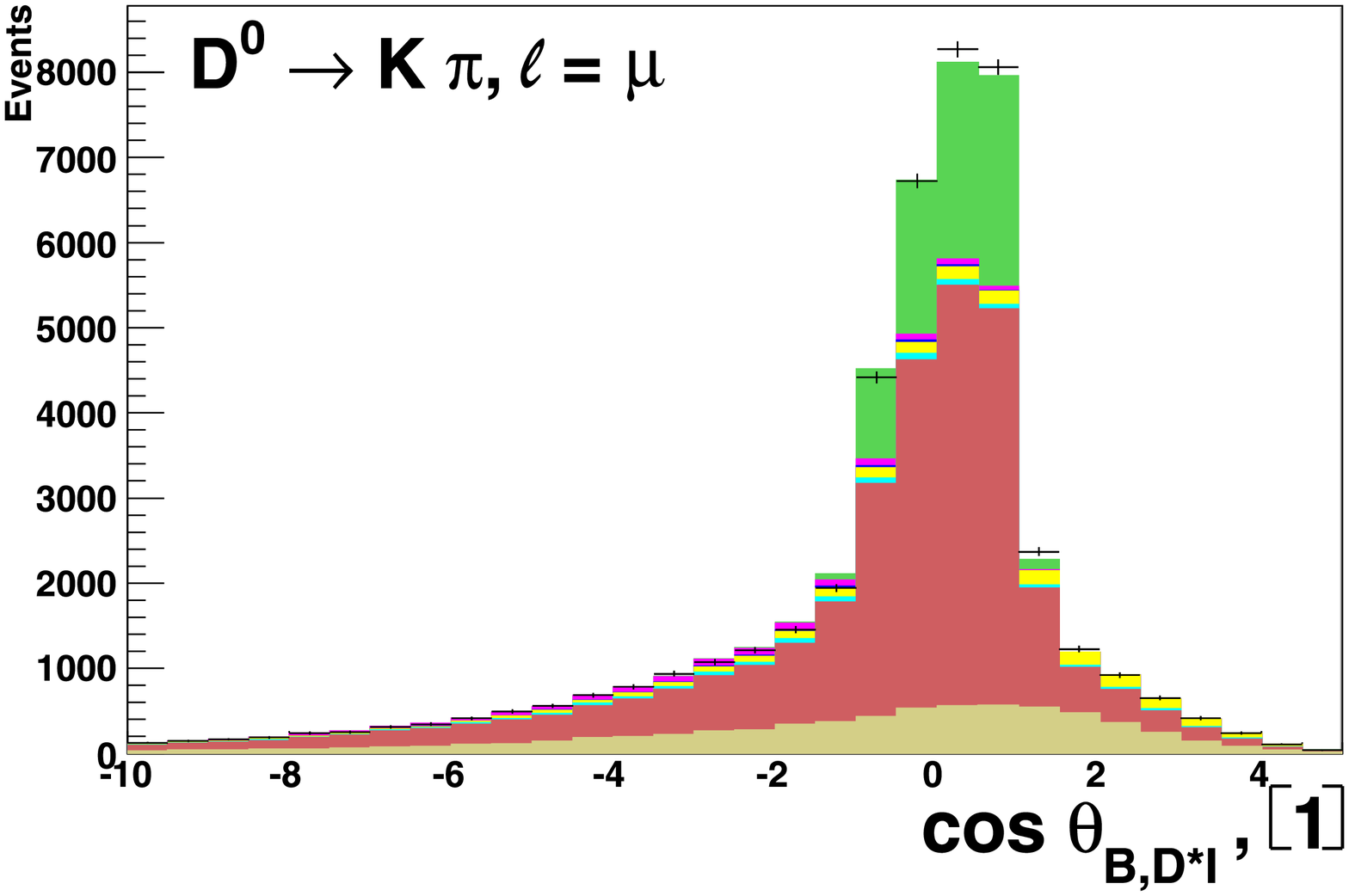}
    \includegraphics[width=0.45\columnwidth]{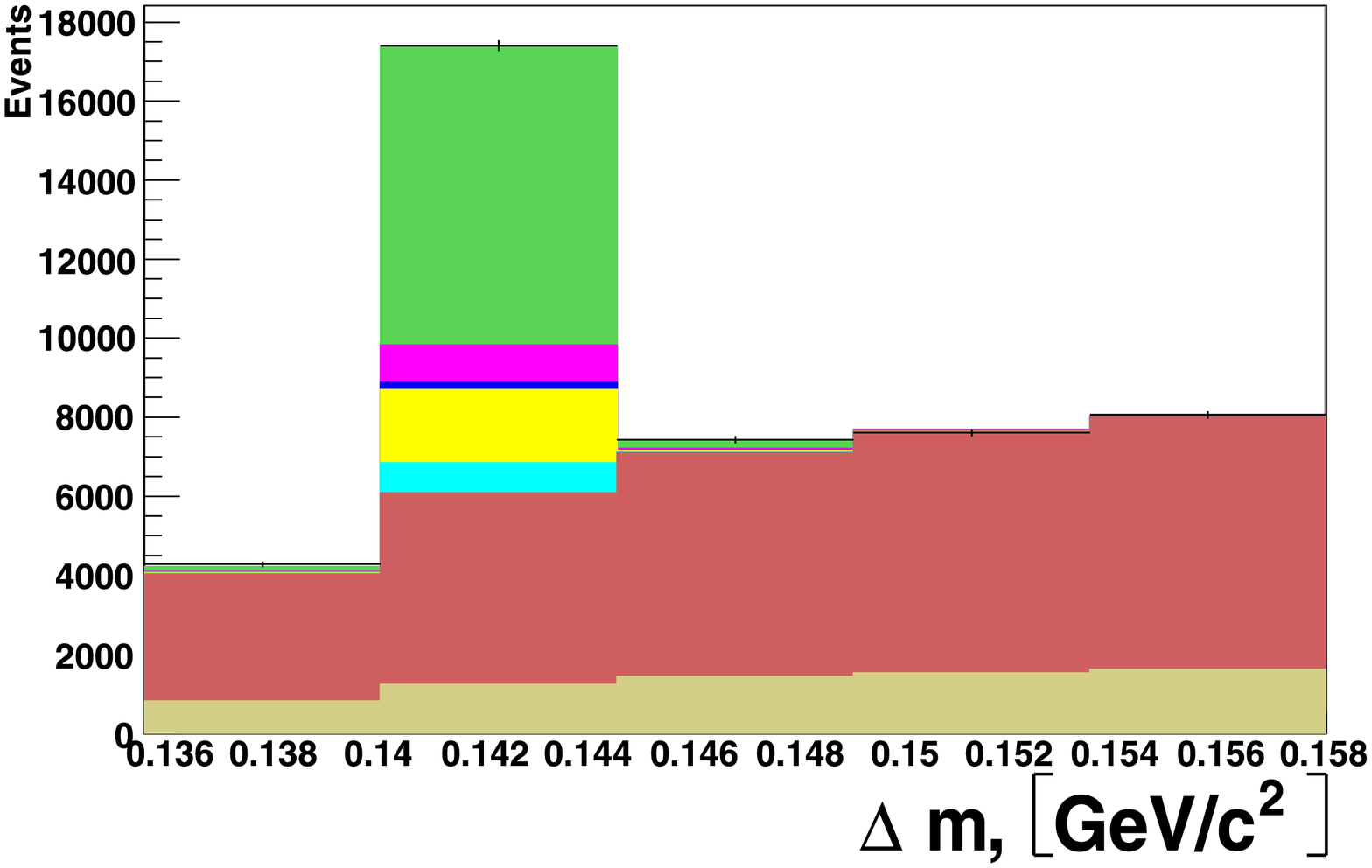}
  \end{center}
  \caption{Same of Fig.~\ref{fig:5a} for the $K\pi, \mu$ sub-sample.}
  \label{fig:5b}
\end{figure}

\begin{figure}
  \begin{center}
    \includegraphics[width=0.45\columnwidth]{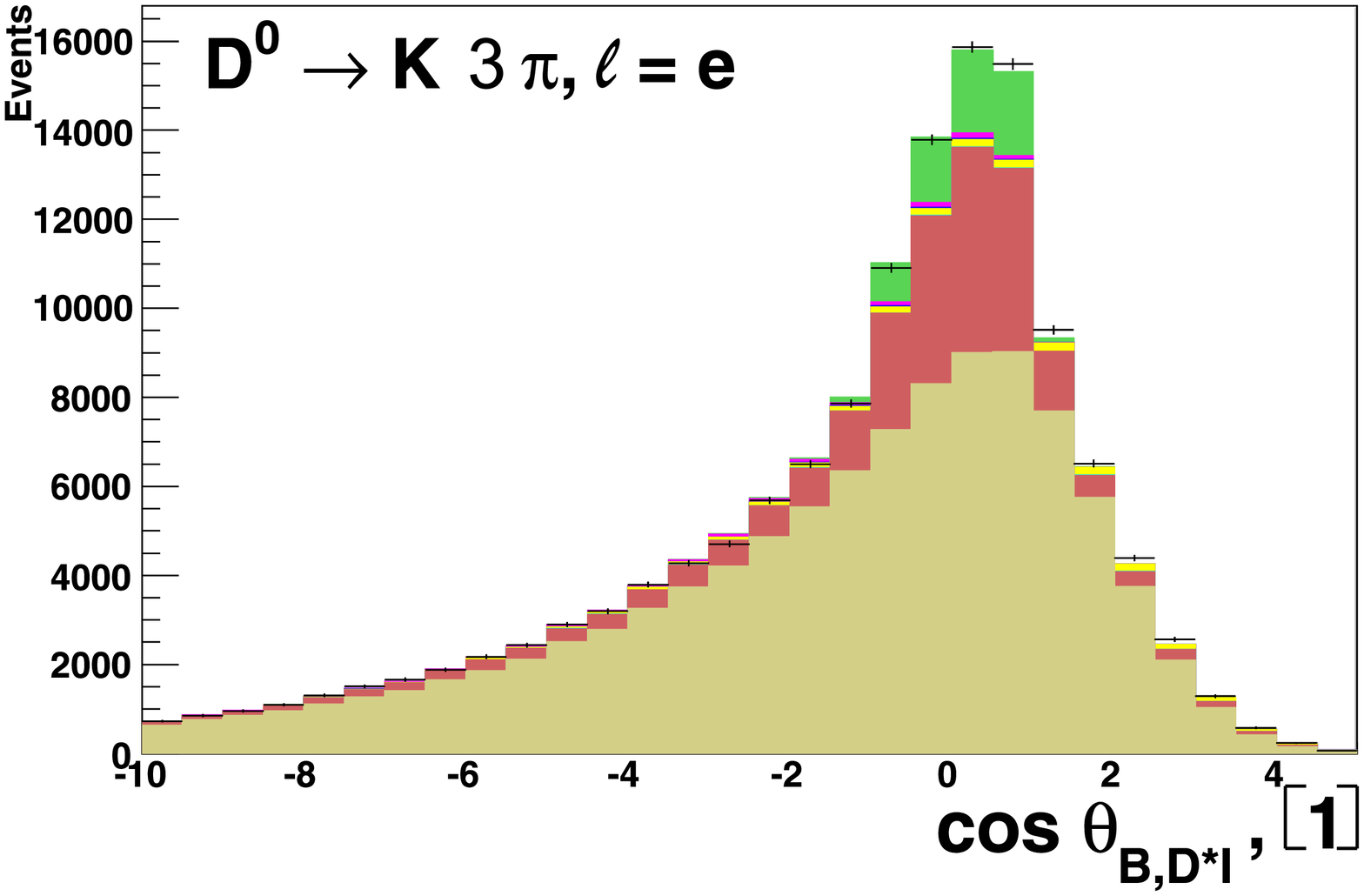}
    \includegraphics[width=0.45\columnwidth]{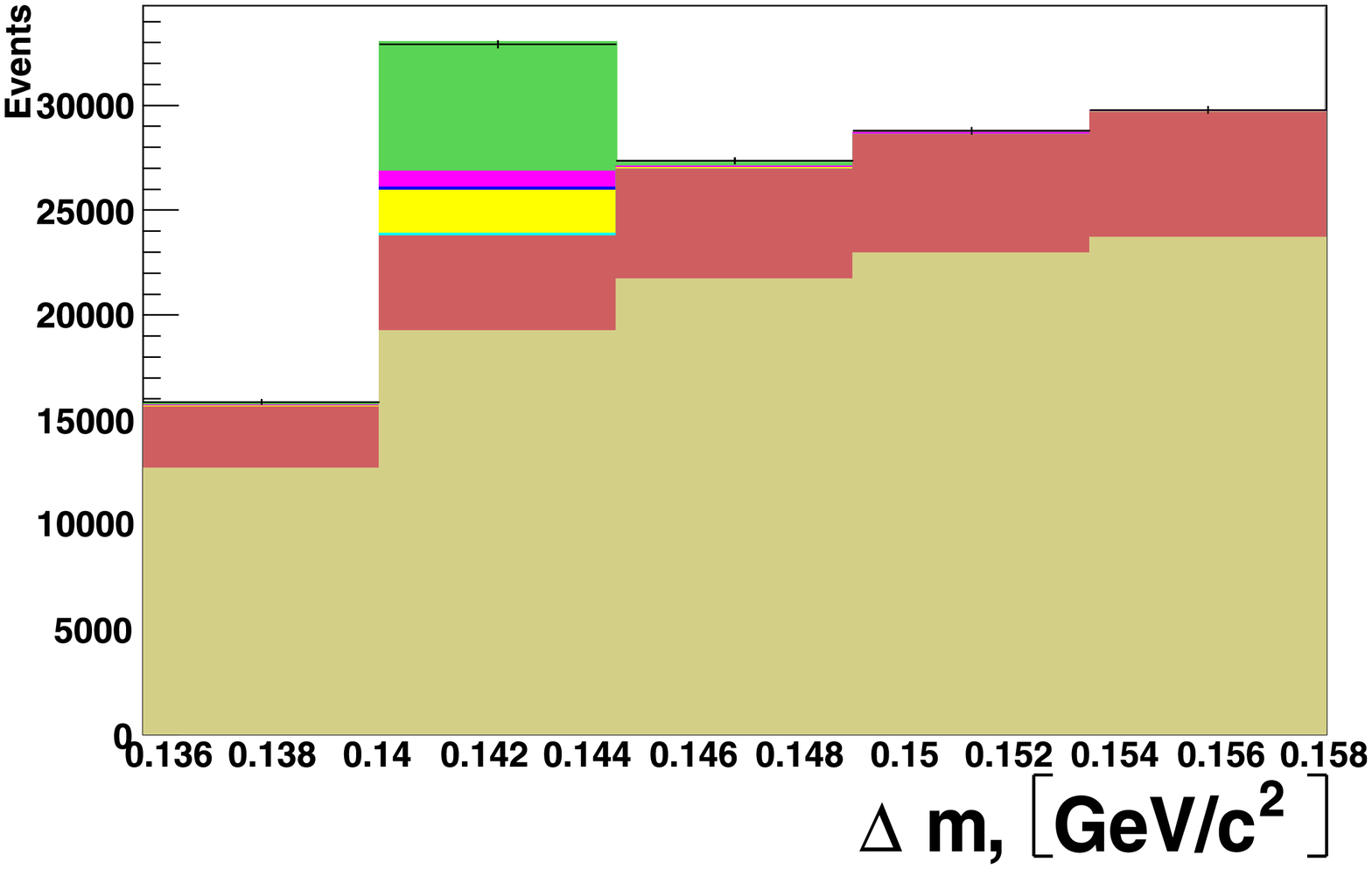}
  \end{center}
  \caption{Same of Fig.~\ref{fig:5a} for the $K3\pi, e$ sub-sample.}
  \label{fig:5c}
\end{figure}

\begin{figure}
  \begin{center}
    \includegraphics[width=0.45\columnwidth]{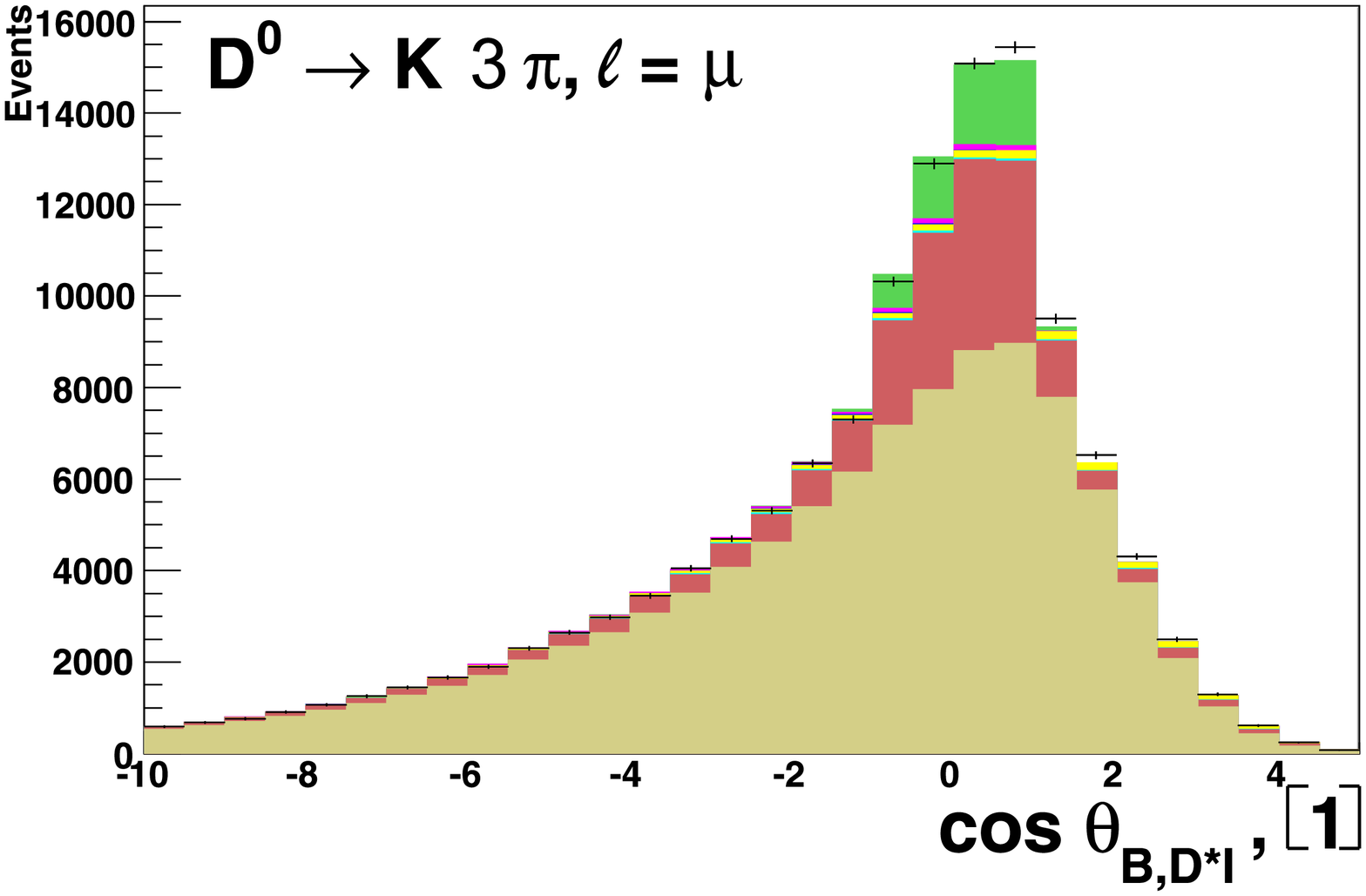}
    \includegraphics[width=0.45\columnwidth]{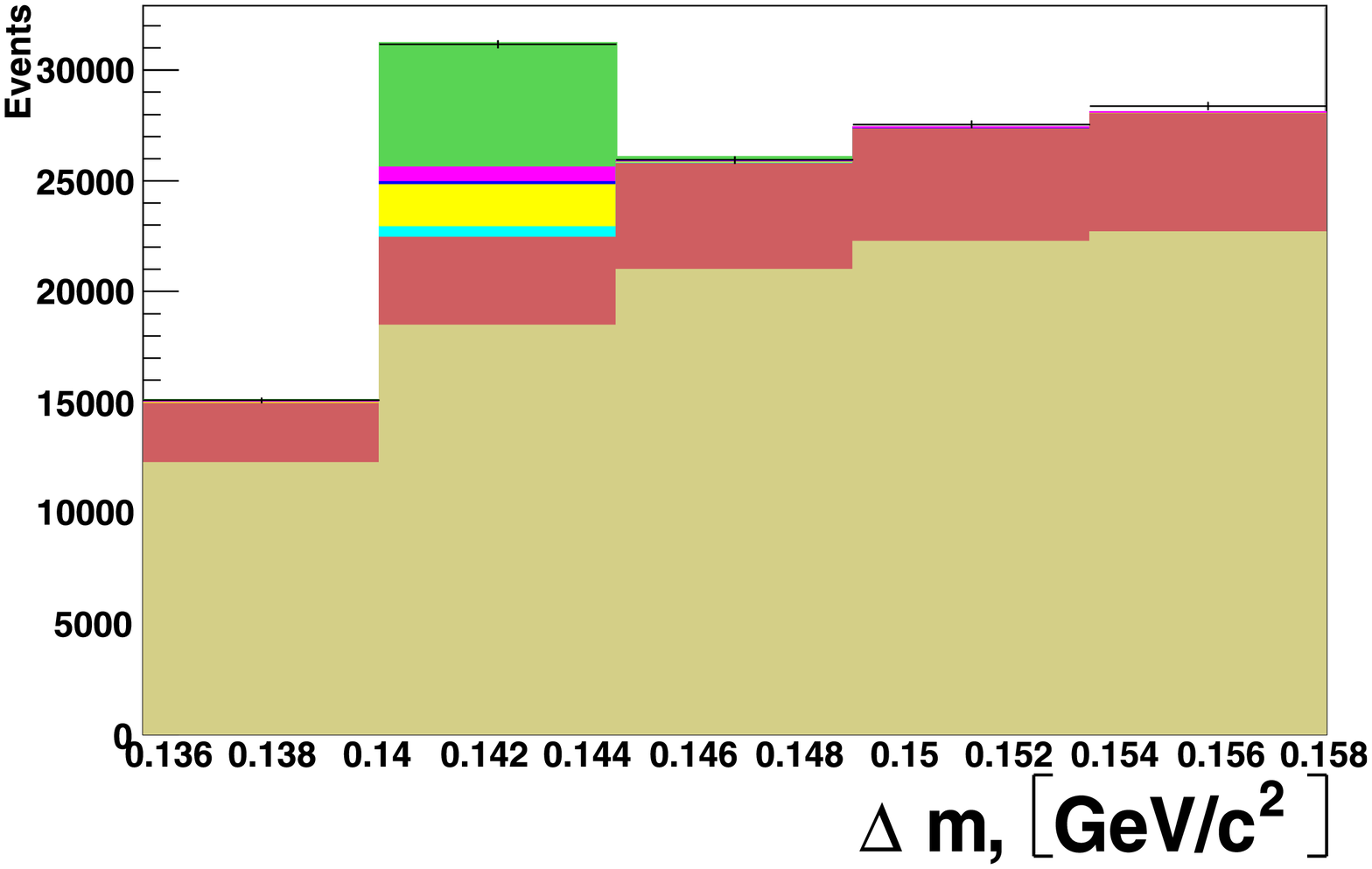}
  \end{center}
  \caption{Same of Fig.~\ref{fig:5a} for the $K3\pi, \mu$ sub-sample.}
  \label{fig:5d}
\end{figure}

These background components are modeled by MC $(D^{*0}\ell)$
events, appropriately selected using generator information, except for
continuum which is modeled by off-resonance events. For muon events,
the shape of the fake lepton background is corrected by the ratio of
the pion fake rate in the experimental data over the same quantity in
the MC. The lepton identification efficiency is reweighted 
to compensate for imperfect MC simulation of the detector. The 
actual fit uses the 
{\tt TFractionFitter} algorithm~\cite{Barlow:1993dm} in {\tt  ROOT}
~\cite{Brun:1997pa}.  The fit is done separately in the four 
sub-samples defined by the $D^0$  decay channel and the lepton type. 
The results are shown in Figs.~\ref{fig:5a} to \ref{fig:5d} and 
Table~\ref{tab:1}.
\begin{table}
  \begin{center}
  	\renewcommand\arraystretch{1.2}
\begin{tabular}{l|@{\extracolsep{.2cm}}cccc}
\hline\hline
           & $K \pi, e$ & $K \pi, \mu$ & $K3\pi, e$ & $K3\pi, \mu$  \\
\hline
Raw yield &      13035 &      12262 &      16989 &      16350  \\
Signal events &      8133  $\pm$       205  &      7447  $\pm$       201  &      5987  $\pm$       229  &      5539  $\pm$       222  \\
\hline
    Signal & (62.39 $\pm$       1.57)\% & (60.73 $\pm$       1.64)\% & (35.24 $\pm$       1.35)\% & (33.88 $\pm$       1.36)\%\\

Signal correlated & (1.27 $\pm$       0.31)\% & (1.46 $\pm$       0.32)\% & (1.16 $\pm$       0.26)\% & (1.34 $\pm$       0.31)\%\\

$D^{**}$ & (0.77 $\pm$       0.98)\% & (0.73 $\pm$       0.98)\% & (0.39 $\pm$       0.50)\% & (0.36 $\pm$       0.47)\% \\

Uncorrelated & (4.97 $\pm$       0.54)\% & (4.25 $\pm$       0.45)\% & (3.48 $\pm$       0.41)\% & (3.30 $\pm$       0.38)\%\\

Fake $\ell$ & (0.31 $\pm$       0.10)\% & (1.94 $\pm$       0.59)\% & (0.18 $\pm$       0.06)\% & (0.95 $\pm$       0.29)\%\\

Combinatoric $D^{*0}$ & (24.76 $\pm$       0.51)\% & (24.30 $\pm$       0.48)\% & (16.35 $\pm$       0.69)\% & (15.19 $\pm$       0.67)\%\\

Fake $D^0$ & (2.91 $\pm$       0.25)\% & (3.12 $\pm$       0.23)\% & (38.53 $\pm$       0.50)\% & (39.45 $\pm$       0.51)\%\\

Continuum & (2.63 $\pm$       0.43)\% & (3.46 $\pm$       0.51)\% & (4.68 $\pm$       0.50)\% & (6.14 $\pm$       0.56)\%\\
\hline\hline
\end{tabular}

  \end{center}
  \caption{The signal and background fractions for selected events
    within the signal window $|\cos\theta_{B,D^{*0}\ell}|<1$ and
    $140~\mathrm{MeV}<\Delta m<144.5~\mathrm{MeV}$.} \label{tab:1}
\end{table}

In all fits, the continuum normalization is fixed to the on- to
off-resonance luminosity ratio, corrected for the $1/s$~dependence of
the $e^+e^-\to q\bar q$ cross-section. The on- to off-resonance luminosity
ratio is measured using Bhabha events to an accuracy of 1.5 \%. 
Aside from the before mentioned corrections to the lepton
identification, the normalization of the fake lepton component has also been
fixed to the MC expectations, since the fit failed to determine this
small background contribution reliably. For the uncertainty in this
component a very conservative value of 30\% is assumed.
The determination of the $D^{**}$ and
signal correlated backgrounds works reliably in the two $D^0\to
K^-\pi^+$~modes, but fails in the three pion channels. In these cases we
fix the data to MC ratio to the value found in the former channels. We
assume the relative uncertainty found in the $K\pi$~modes.

In Table~\ref{tab:1} those components which have been fixed in the fit are
shown with the corresponding error estimate described above. The error
of the continuum component is appropriately scaled by the on- to
off-resonance luminosity ratio.

In the rest of this analysis, we require $D^{*0}\ell$~candidates to
pass the signal window requirement $-1 <\cos\theta_{B,D^{*0}\ell}<1$
and $140~\mathrm{MeV}<\Delta m<144.5~\mathrm{MeV}$.

\subsection{Kinematic variables}

\begin{figure}
  \begin{center}
    \includegraphics[width=0.8\columnwidth]{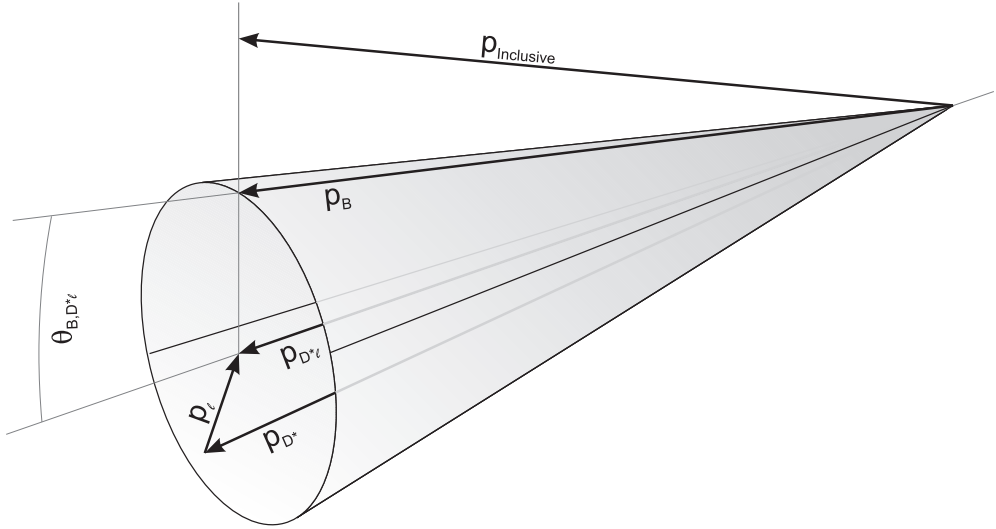}
  \end{center}
  \caption{Reconstruction of the $B^+$~direction. Refer to the text
    for details.} \label{fig:6}
\end{figure}

To calculate the four kinematic variables -- $w$, $\cos\theta_\ell$,
$\cos\theta_V$ and $\chi$ -- that characterize the $B^+\to\bar
D^{*0}l^+\nu$~decay defined in Sect.~\ref{sec:2a}, we need to
determine the $B$~rest frame. The $B$~direction is already known to be
on a cone around the $(D^{*0}\ell)$-axis with opening angle
$2\theta_{B,D^{*0}\ell}$ in the c.m.\ frame, Eq.~\ref{eq:3_1}. For the
best determination of the $B$~direction, we first estimate the c.m.\ frame
$B$~vector by summing the momenta of the remaining particles in the
event ($\vec p^*_\mathrm{inclusive}$~\cite{ref:2}) and choose the
direction on the cone that minimizes the difference to $\vec
p^*_\mathrm{inclusive}$, Fig.~\ref{fig:6}.

To obtain $\vec p^*_\mathrm{inclusive}$, we first apply the following
track and cluster selection requirements to particles which have not
been associated with the $(D^{*0}\ell)$~system: neutral clusters are
required to satisfy $E_\gamma>100$~MeV for polar angles with respect
to the beam direction $\theta<32^{\circ}$, $E_\gamma>150$~MeV for
$\theta>130^{\circ}$ and $E_{\gamma}>50$~MeV in the barrel region in
between; the impact parameter cuts for charged tracks are $d
r<20$~cm and $|d z|<100$~cm for $p_\mathrm{T}<0.25$~GeV/$c$,
$d r<15$~cm and $|d z|<50$~cm for $p_\mathrm{T}<0.5$~GeV/$c$,
and $d r< 10$~cm and $|d z|<20$~cm for
$p_\mathrm{T}>0.5$~GeV/$c$. We exclude duplicated tracks by selecting
pairs of charged tracks with $p_\mathrm{T}<275$~MeV/$c$, momentum
difference~$\Delta p<100$~MeV/$c$ and relative angle smaller than
15$^\circ$ or larger than 165$^\circ$. If such a pair is found, only
the track with smaller value of $(5~dr)^2+(|d z|)^2$ is
retained.

Then, we compute $\vec p_\mathrm{inclusive}$ (in the lab.\ frame) by
summing the 3-momenta of the selected particles,
\begin{equation}
  \vec p_\mathrm{inclusive}=\vec p_\mathrm{HER}+\vec
  p_\mathrm{LER}-\sum_i \vec p_i~,
\end{equation}
where the indices HER and LER correspond to the colliding beams, and
transform this vector into the c.m.\ frame. Note that we do not make
any mass assumption for the charged particles. The energy component of
the 4-vector $P_\mathrm{inclusive}$ is defined by requiring
$E^*_\mathrm{inclusive}$ to be $E^*_\mathrm{beam}=\sqrt{s}/2$.

With the $B$~rest frame reconstructed in this way, the resolutions
in the kinematic variables are found to be about 0.021, 0.049, 0.057
and 6.46$^\circ$ for $w$, $\cos\theta_\ell$, $\cos\theta_V$ and
$\chi$, respectively.

\subsection{Fit procedure} \label{section:exp-parametrizedFit}

We perform a binned $\chi^2$~fit of the $w$, $\cos\theta_\ell$,
$\cos\theta_V$ and $\chi$~distributions over (almost) the entire phase
space to measure the following quantities: the form factor
normalization $\mathcal{F}(1)|V_{cb}|$, Eq.~\ref{eq:2_2}, and the
three parameters $\rho^2$, $R_1(1)$ and $R_2(1)$ which parameterize
the form factor in the HQET framework,
Eqs.~\ref{eq:2_3}--\ref{eq:2_5}. Instead of fitting in four
dimensions, we fit the one-dimensional projections of $w$,
$\cos\theta_\ell$, $\cos\theta_V$ and $\chi$ to have enough entries in
each bin of the fit. This introduces bin-to-bin correlations which
have to be accounted for.

The distributions in $w$, $\cos\theta_\ell$, $\cos\theta_V$ and $\chi$
are divided into ten bins of equal width. The kinematically allowed
values of $w$ are between $1$ and $\approx 1.505$ but we restrict the
$w$~range to values between $1$ and $1.5$. In each sub-sample, there
are thus 40~bins to be used in the fit. In the following, we label
these bins with a common index $i$, $i=1,\dots,40$. The bins $i=1,\dots,10$
correspond to the bins of the $w$~distribution, $i=11,\dots,20$ to
$\cos\theta_\ell$, $i=21,\dots,30$ to $\cos\theta_V$ and $i=31,\dots,40$ to
the $\chi$ distribution.

The predicted number of events $N^\mathrm{th}_i$ in the bin~$i$ is
given by
\begin{equation}
  N^\mathrm{th}_i=N_{B^+}\mathcal{B}(D^{*0}\to
  D^0\pi^0)\mathcal{B}(D^0)\tau_{B^+}\Gamma_i~,
\end{equation}
where $N_{B^+}$ is the number of $B^+$~mesons in the data sample and
$\mathcal{B}(D^{*0}\to D^0\pi^0)$ is taken from
Ref.~\cite{Amsler:2008zzb}. In the $(K\pi,e)$ and $(K\pi,\mu)$
sub-samples $\mathcal{B}(D^0)$ is $\mathcal{B}(D^0\to
K^-\pi^+)$~\cite{Amsler:2008zzb}; in $(K3\pi,e)$ and $(K3\pi,\mu)$
$\mathcal{B}(D^0)$ is $R_{K3\pi/K\pi}\mathcal{B}(D^0\to K^-\pi^+)$,
with $R_{K3\pi/K\pi}$ a fifth free parameter of the fit. Finally,
$\tau_{B^+}$ is the $B^+$~lifetime~\cite{Amsler:2008zzb}, and
$\Gamma_i$ is the width obtained by integrating Eq.~\ref{eq:2_1} in
the kinematic variable corresponding to $i$ from the lower to the
upper bin boundary (the other kinematic variables are integrated over
their full range). This integration is numerical in the case of $w$
and analytic for the other variables. The expected number of events
$N^\mathrm{exp}_i$ is related to $N^\mathrm{th}_i$ as follows
\begin{equation}
  N^\mathrm{exp}_i=\sum_{j=1}^{40}\left(R_{ij}\epsilon_jN^\mathrm{th}_j\right)+N^\mathrm{bkgrd}_i~.
  \label{eq:ApplicationOfEfficiencies}
\end{equation}
Here, $\epsilon_i$ is the probability that an event generated in the
bin~$i$ is reconstructed and passes all analysis cuts and $R_{ij}$ is the
detector response matrix, \textit{i.e.}, it gives the probability that
an event generated in the bin~$j$ is observed in the bin~$i$. Both
quantities are calculated using MC~simulation. $N^\mathrm{bkgrd}_i$ is
the number of expected background events, estimated as described in
Sect.~\ref{sec:3c}.

Next, we calculate the variance~$\sigma^2_i$ of $N^\mathrm{exp}_i$. We
consider the following contributions: the poissonian uncertainty in
$N^\mathrm{th}_i$; fluctuations related to the efficiency, which are 
investigated using a binomial distribution with $N$ repetitions and the
success probability $\epsilon_i$; a similar contribution based on a 
multinomial distribution related to $R_{ij}$ ; and the uncertainty in the
background contribution $N^\mathrm{bkgrd}_i$. This yields the
following expression for $\sigma^2_i$,
\begin{equation}
  \begin{split}
    \sigma^2_i=\sum_{j=1}^{40}\left[R^2_{ij}\epsilon^2_jN^\mathrm{th}_j+R^2_{ij}\frac{\epsilon_j(1-\epsilon_j)}{N_\mathrm{data}}(N^\mathrm{th}_j)^2+\frac{R_{ij}(1-R_{ij})}{N'_\mathrm{data}}\epsilon^2_j(N^\mathrm{th}_j)^2+\right.\\
      \left.R^2_{ij}\frac{\epsilon_j(1-\epsilon_j)}{N_\mathrm{MC}}(N^\mathrm{th}_j)^2+\frac{R_{ij}(1-R_{ij})}{N'_\mathrm{MC}}\epsilon^2_j(N^\mathrm{th}_j)^2\right]+\sigma^2(N^\mathrm{bkgrd}_i)~.
      \label{eq:sigmaErrorPropagation}
  \end{split}
\end{equation}
The first term is the poissonian uncertainty in $N^\mathrm{th}_i$. The
second and third terms are the binomial respectively multinomial uncertainties
related to the finite real data size, where $N_\mathrm{data}$
($N'_\mathrm{data}$) is the total number of decays (the number of
reconstructed decays) into the final state under consideration ($K\pi$
or $K3\pi$, $e$ or $\mu$) in the real data. The quantities $\epsilon_i$
and $R_{ij}$ are calculated from a finite signal MC sample
($N_\mathrm{MC}$ and $N'_\mathrm{MC}$); the corresponding
uncertainties are estimated by the fourth and fifth terms. Finally,
the last term is the background contribution
$\sigma^2(N^\mathrm{bkgrd}_i)$, calculated as the sum of the different
background component variances. For each background component
defined in Sect.~\ref{sec:3c} we estimate its contribution by linear error
propagation of the scale factor and the error determined by the
procedure described above.

In each sub-sample we calculate the off-diagonal elements of the
covariance matrix~cov$_{ij}$ as $Np_{ij}-Np_ip_j$, where $p_{ij}$ is
the relative abundance of the bin~$(i,j)$ in the 2-dimensional histogram
obtained by plotting the kinematic variables against each other,
$p_i$ is the relative number of entries in the 1-dimensional
distribution, and $N$ is the size of the sample. Covariances are
calculated for the signal and the different background components, and
added with appropriate normalizations.

The covariance matrix is inverted numerically within {\tt
  ROOT} and, labelling the four sub-samples ($K\pi$
or $K3\pi$, $e$ or $\mu$) with the index~$k$, the sub-sample
$\chi^2$~functions are calculated,
\begin{equation}
  \chi^2_k=\sum_{i,j}(N^\mathrm{obs}_i-N^\mathrm{exp}_i)C^{-1}_{ij}(N^\mathrm{obs}_j-N^\mathrm{exp}_j)~,
\end{equation}
where $N^\mathrm{obs}_i$ is the number of events observed in bin~$i$
in the data. We sum these four functions and minimize the global
$\chi^2$ with MINUIT~\cite{James:1975dr}.


We have tested this fit procedure for possible biases using MC samples, 
generated both with and without full detector simulation. All results 
are consistent with expectations and show no indication of bias.


\section{Results and systematic uncertainties}

\subsection{Results}

\begin{figure}
  \begin{center}
    \includegraphics[width=0.8\columnwidth]{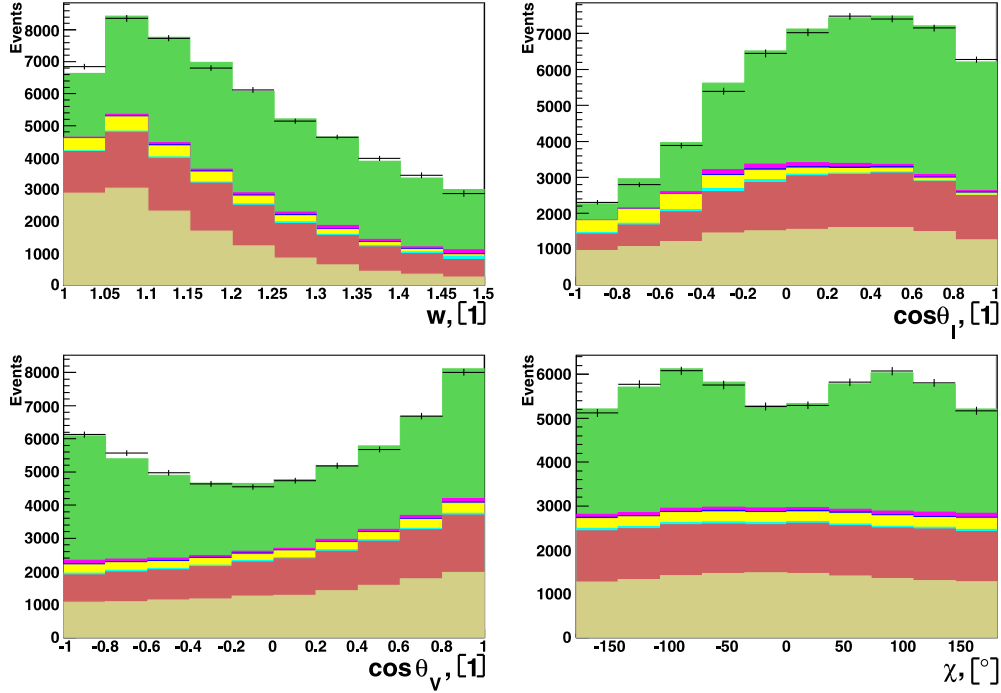}
  \end{center}
  \caption{Result of the fit of the four kinematic variables in the
    total sample. (The different sub-samples are added in this plot.)
    The points with error bars are continuum subtracted on-resonance
    data. The histograms are the signal and the different background
    components. The color scheme is explained in Fig.~\ref{fig:3}.}
    \label{fig:7}
\end{figure}

After applying all analysis cuts and subtracting backgrounds,
$27,106\pm 367$ signal events are found in the data. The preliminary
result of the fit to these events is shown in Fig.~\ref{fig:7} and
Tab.~\ref{tab:4}. The correlation coefficients of the five fit
parameters are given in Tab.~\ref{tab:5}. The breakup of the
systematic uncertainty is given in Tab.~\ref{tab:6}.
\begin{table}
  \begin{center}
\begin{tabular}{l|@{\extracolsep{.2cm}}ccc}
\hline\hline
           & $D^0\to K\pi$.$\ell= e$ & $D^0\to K\pi$. $\ell=\mu$ & $D^0\to K3\pi$. $\ell=e$\\
\hline
 $\rho^2$ &     1.199  $\pm$     0.125  $\pm$     0.048  &     1.370  $\pm$     0.129  $\pm$     0.057  &     1.723  $\pm$     0.162  $\pm$     0.061  \\
 $R_1(1)$ &     1.507  $\pm$     0.135  $\pm$     0.090  &     1.568  $\pm$     0.158  $\pm$     0.088  &     1.840  $\pm$     0.271  $\pm$     0.109  \\
 $R_2(1)$ &     0.868  $\pm$     0.093  $\pm$     0.034  &     0.839  $\pm$     0.110  $\pm$     0.031  &     0.585  $\pm$     0.198  $\pm$     0.047  \\
$R_{K3\pi/K\pi}$ &     2.072  &     2.072  &     2.072    \\
$\mathcal{B}(B^+\to\bar D^{*0}\ell^+\nu_\ell)$ &      4.91  $\pm$      0.05  $\pm$      0.57  &      4.77  $\pm$      0.05  $\pm$      0.56  &      4.83  $\pm$      0.07  $\pm$      0.56  \\
$\mathcal{F}(1)|V_{cb}|\times 10^3$ &      34.3  $\pm$       0.6  $\pm$       2.2  &      35.0  $\pm$       0.6  $\pm$       2.3  &      36.5  $\pm$       1.0  $\pm$       2.3  \\
$\chi^2/$ndf.\ &      48.3  /        36  &      40.6  /        36  &      39.6  /        36              \\
$P_{\chi^2}$ &       8.3  \%            &      27.5  \%            &      31.3  \%            \\
\hline\hline
\end{tabular}  

\vspace{.3cm}
    
\begin{tabular}{l|@{\extracolsep{.2cm}}ccc}
\hline\hline
           & $D^0\to K3\pi$. $\ell=\mu$ &  & Fit to total sample \\
\hline
 $\rho^2$ &     1.434  $\pm$     0.209  $\pm$     0.086  &            &1.376  $\pm$     0.074  $\pm$     0.056  \\
 $R_1(1)$ &     1.813  $\pm$     0.273  $\pm$     0.106  &            &1.620  $\pm$     0.091  $\pm$     0.092  \\
 $R_2(1)$ &     0.764  $\pm$     0.191  $\pm$     0.051  &            &0.805  $\pm$     0.064  $\pm$     0.036  \\
$R_{K3\pi/K\pi}$ &     2.072  &            & 2.072  $\pm$     0.023              \\
$\mathcal{B}(B^+\to\bar D^{*0}\ell^+\nu_\ell)$ &      4.83  $\pm$      0.07  $\pm$      0.57  &            &      4.84  $\pm$      0.04  $\pm$      0.56  \\
$\mathcal{F}(1)|V_{cb}|\times 10^3$ &      34.8  $\pm$       1.0  $\pm$       2.3  &            &      35.0  $\pm$       0.4  $\pm$       2.2  \\
$\chi^2/$ndf.\ &      44.2  /        36  &            &     187.8  /       155              \\

$P_{\chi^2}$ &      16.3  \%            &  \phantom{1.434  $\pm$     0.209  $\pm$     0.086}    &             3.7  \%     \\  
\hline\hline
\end{tabular}
  \end{center}
  \caption{The results of the fits to the sub-samples and to the total
  sample. The first error is statistical, the second is the estimated
  systematic uncertainty.
  } 
  \label{tab:4}
\end{table}
\begin{table}
  \begin{center}
  	\begin{tabular}{l|@{\extracolsep{.2cm}}cccc}

\hline\hline
 & $\mathcal{F}(1)|V_{cb}|$ &  $\rho^2$ &  $R_1(1)$ &  $R_2(1)$  \\
\hline
$\mathcal{F}(1)|V_{cb}|$ &	1.000 &	0.455/0.399/0.295 &	-0.222 /-0.219/-0.179 &	-0.054/-0.024/-0.019 \\
$\rho^2$	&		&			1.000 	& 	\phantom{-}0.648/\phantom{-}0.413/\phantom{-}0.540 &	-0.889/-0.751/-0.841 \\
$R_1(1)$	&		&   &																		1.000 &																-0.749/-0.873/-0.763 \\
$R_2(1)$	&		&		&		&																																								1.000 \\
\hline\hline
\end{tabular}
  \end{center}
  \caption{The statistical/systematic/total correlation coefficients 
  for $\mathcal{F}(1)|V_{cb}|$ and the three form factor parameters in
  the fit to the full sample.} \label{tab:5}
\end{table}

As explained earlier, the branching fraction of the decay $D^0\to
K^-\pi^+\pi^+\pi^-$ is floated in the fit to the full sample. The fit
result is in good agreement with the world
average~\cite{Amsler:2008zzb}. In the sub-sample fits, $R_{K3\pi/K\pi}$
is fixed to the value of the full sample fit.

\subsection{Systematic uncertainties}

To estimate the systematic errors in the results quoted above,
we consider uncertainties in the following: signal reconstruction
efficiency, background estimation, $D^{*0}\to D^0\pi^0$ and $D^0\to
K^-\pi^+$ branching fractions~\cite{Amsler:2008zzb},
$B^+$~lifetime~\cite{Amsler:2008zzb}, and the number of $B^+$~mesons
in the data sample.

To calculate these systematic uncertainties, we consider 300
pseudo-experiments in which 13 quantities (corresponding to the
above-mentioned contributions) are randomly varied, taking into account
possible correlations. The entire analysis chain is repeated for every
pseudo-experiment and new fit values are obtained. One standard
deviation of the pseudo-experiment fit results for a given parameter
is used as the systematic uncertainty in this parameter. This toy MC
approach also allows the systematic correlation coefficients
in Table~\ref{tab:5} to be derived in a straightforward way.

In the following, we describe the parameters varied in the
pseudo-experiments:
\begin{itemize}
\item The tracking efficiencies in five bins of slow pion momentum are
  varied within their respective uncertainties. These uncertainties
  are determined from real data by studying the decay $B^+\to\bar
  D^{*0}\pi^+$ in the same 140~fb$^{-1}$ sample as used for the main
  analysis. The tracking uncertainties in different momentum bins are
  considered fully correlated. Therefore, tracking corresponds only to
  a single parameter in the toy MC,

\item The lepton identification uncertainties are varied within their
  respective uncertainties~\cite{Hanagaki:2001fz,Abashian:2002bd},

\item The normalizations of the 7 background components are varied
  within the uncertainties determined by the background fit in
  Sect.~\ref{sec:3c}. For continuum, the on- to off-resonance
  luminosity ratio is varied within its uncertainty of 1.5\%. 
  Correlations between different background component 
  normalizations are taken into account,

\item Additionally, we vary the shape in $w$ of the uncorrelated,
  combinatoric $D^{*0}$ and fake $D^0$ components. We fit each of
  these distributions by a Lorentz distribution and vary each of the
  parameters within the errors obtained in the fit,
  
\item We vary the fraction $f_{+-}/f_{00}=\mathcal{B}(\Upsilon(4S)\to
  B^+B^-)/\mathcal{B}(\Upsilon(4S)\to B^0\bar B^0)$ within its
  uncertainty~\cite{Amsler:2008zzb}. This parameter affects the
  background distributions.

\end{itemize}

The uncertainties in $\mathcal{B}(D^{*0}\to D^0\pi^0)$,
$\mathcal{B}(D^0\to K^-\pi^+)$, the number of $B^+$ in the sample and
the $B^+$~lifetime affect only $\mathcal{F}(1)|V_{cb}|$, not the form
factors. Therefore, they are not considered in the toy MC.

The breakup of the systematic errors quoted in Table~\ref{tab:4} is
given in Table~\ref{tab:6}.
\begin{table}
  \begin{center}
  	\renewcommand\arraystretch{1.2}
\begin{tabular}{l|@{\extracolsep{.2cm}}ccccc}
\hline\hline
           &  $\rho^2$ &  $R_1(1)$ &  $R_2(1)$ & $\mathcal{F}(1)|V_{cb}|\times 10^3$ & $\mathcal{B}(B^+\to\bar D^{*0}\ell^+\nu_\ell)$ \\
\hline
     Value &     1.376  &     1.620  &     0.805  &     34.98  &     4.841  \\

Statistical Error &     0.074  &     0.091  &     0.064  &      0.37  &     0.044  \\
\hline
  Tracking &     -0.027  &     +0.025  &     +0.012  &      -1.97  &     -0.491  \\

  LeptonID &     +0.012  &     +0.024  &     -0.011  &      -0.39  &     -0.096  \\

Norm - Signal Corr. &     -0.007  &     +0.002  &     +0.007  &      +0.13  &     +0.038  \\

Norm - $D^{**}$ &     +0.005  &     -0.023  &     +0.002  &      -0.04  &     -0.041  \\

Norm - Uncorr &     +0.014  &     +0.074  &     -0.025  &      -0.28  &     -0.023  \\

Norm - Fake $\ell$ &     +0.017  &     +0.028  &     -0.010  &      -0.05  &     -0.024  \\

Norm - Comb $D^{*0}$ &     +0.008  &     +0.014  &     -0.008  &      -0.11  &     -0.028  \\

Norm - Fake $D^0$ &     -0.009  &     -0.014  &     +0.007  &      +0.06  &     +0.020  \\

Norm -  Continuum &     +0.004  &    +0.005  &     -0.001  &      0.00  &     -0.003  \\

Shape - Uncorr &     +0.014  &     +0.003  &     -0.005  &      +0.10  &            \\

Shape - Comb $D^{*0}$ &     +0.027  &     -0.005  &     -0.008  &      +0.21  &            \\

Shape - Fake $D^0$ &     +0.024  &     +0.003  &     +0.008  &      +0.17  &            \\

$\mathcal{B}(D^0\to K\pi)$ &            &            &            &      -0.32  &     -0.089  \\

$\mathcal{B}(D^{*0}\to D^0\pi^0)$ &            &            &            &      -0.82  &     -0.227  \\

$B^+$ lifetime &            &            &            &      -0.12  &     -0.033  \\

$N (\Upsilon(4S))$ &            &            &            &      -0.14  &     -0.040  \\

$f_{+-} / f_{00}$ &     +0.003  &     +0.006  &     -0.003  &      -0.15  &     -0.043  \\
\hline\hline
\end{tabular}
	\end{center}
  \caption{The breakup of the systematic uncertainty in the result
  of the fit to the full sample. The sign + (-) implies whether the fit
  result moves to larger (smaller) values, if the value of the corresponding
  systematic parameter is increased.} \label{tab:6}
\end{table}

\section{Model-independent determination of helicity amplitudes}

The results discussed in the previous section rely on the form factor
parameterization by Caprini {\it et al.}~\cite{Caprini:1997mu} and it is
worthwhile to check this assumption. In this section, we extract the
form factor shape through a fit to the $w$ vs.\ $\cos\theta_V$
distribution. For the former quantity, we consider six bins in the
range from $1$ to $1.5$. For the latter, we have also six bins between
$-1$ and $1$. The contribution from events with $w>1.5$ is fixed to
the very small values predicted by the results of the parametrized fit.

\subsection{Fit procedure}

From Eq.~\ref{eq:2_1} we can obtain the double differential decay
width $\mathrm{d}^2\Gamma/\mathrm{d}w\,\mathrm{d}\cos\theta_V$ by
integration over $\cos\theta_\ell$ and $\chi$.

If we define
\begin{equation}
  F_\Gamma = \frac{3 \, G_F^2\, (m_{B}-m_{D^*})^2 \, m_{D^*}^3 }{4^5 \pi^4}
\end{equation}
and
\begin{eqnarray}
  &&\gamma^{\pm\pm}(w) = 
  \sqrt{w^2-1}(w+1)^2h_{A_1}^2(w)|V_{cb}|^2\frac{ 1-2wr-r^2 }{(1-r)^2} \left\{ 1\mp\sqrt{\frac{w-1}{w+1}} R_1(w)\right\}^2,
  \nonumber
  \\
  &&\gamma^{00}(w) =
  \sqrt{w^2-1}(w+1)^2 h_{A_1}^2(w) |V_{cb}|^2 \left\{1 + \frac{w-1}{1-r} \left( 1- R_2(w) \right) \right\}^2,
\end{eqnarray}
the double differential width becomes
\begin{eqnarray}
  &&\frac{\mathrm{d}^2\Gamma(B^+\to\bar D^{*0}\ell^+\nu_\ell)}{\mathrm{d}w\,\mathrm{d}(\cos\theta_V)}
  = \frac{16\pi}{3}F_\Gamma\left(\sin^2 \theta_V \left( \gamma^{++} + \gamma^{--} \right) + 2\,\cos^2\theta_V \, \gamma^{00} \right).
  \label{eq:wVsThetaV}
\end{eqnarray}
The quantities $\gamma^{kk}$ correspond to the $w$-dependence of the
different helicity combinations, times kinematic factors. The one
dimensional distribution, as given in Eq.~\ref{eq:2_2}, depends only
on the sum of these three combinations,
\begin{equation}
  \frac{\mathrm{d}\Gamma(B^+\to\bar D^{*0}\ell^+\nu_\ell)}{\mathrm{d}w} = \frac{64\pi}{9}F_\Gamma\left(\gamma^{++} + \gamma^{--} + \gamma^{00}\right).
\end{equation}

The bin contents of the two dimensional histogram in $w$ vs.\
$\cos\theta_V$ can be obtained by integration of
Eq.~\ref{eq:wVsThetaV} over the corresponding bin area and considering
the reconstruction efficiencies and detector response as described in
Eq.~\ref{eq:ApplicationOfEfficiencies}. Each bin content can be given
as the linear combination of two linearly independent parts. The
integration of the angular distributions is executed analytically,
as described in Sect.~\ref{section:exp-parametrizedFit}, the
integration with respect to $w$ defines a set of free parameters,
\begin{equation}
  \Gamma_{i}^{kk} = \int\limits_{w_i}^{w_{i+1}} \mathrm{d} w \; \gamma^{kk},
\end{equation}
where $w_j = \{ w_1, w_2, \dots, w_7 \} = \{1, 13/12, \dots, 1.5 \}$
are the bin boundaries of the $6$ bins in $w$. Additionally we define
$\gamma^{T} = \gamma^{++} + \gamma^{--}$ and $\Gamma_{i}^{T} =
\Gamma_{i}^{++} + \Gamma_{i}^{--}$.

We define the $\chi^2$ function
$
  \tilde{\chi}^2_{w,\theta_V} = \sum_{i=1}^{6}\sum_{j=1}^{6} \left(\frac{N^{obs}_{ij} - N^{exp}}{\sigma_{N^{exp}} }\right)^2,
$
which depends only on the parameters $\Gamma_{i}^{T}$ and
$\Gamma_{i}^{00}$. Here $N^{obs}$ gives the number of events observed
in on-resonance data, $N^{exp}$ the number of expected events, as
defined in Eq.~\ref{eq:ApplicationOfEfficiencies}, and
$\sigma_{N^{exp}}$ the uncertainty in the expected number of events,
as given in Eq.~\ref{eq:sigmaErrorPropagation}. This expression is
minimized numerically using MINUIT to determine the partial integrals.



Investigation of dedicated sets of toy Monte Carlo showed the
reliability of the procedure described above. The
results show no indication of bias in either the mean or the errors.



\subsection{Results}

\begin{figure}
  \begin{center}
    \includegraphics[width=0.4\columnwidth]{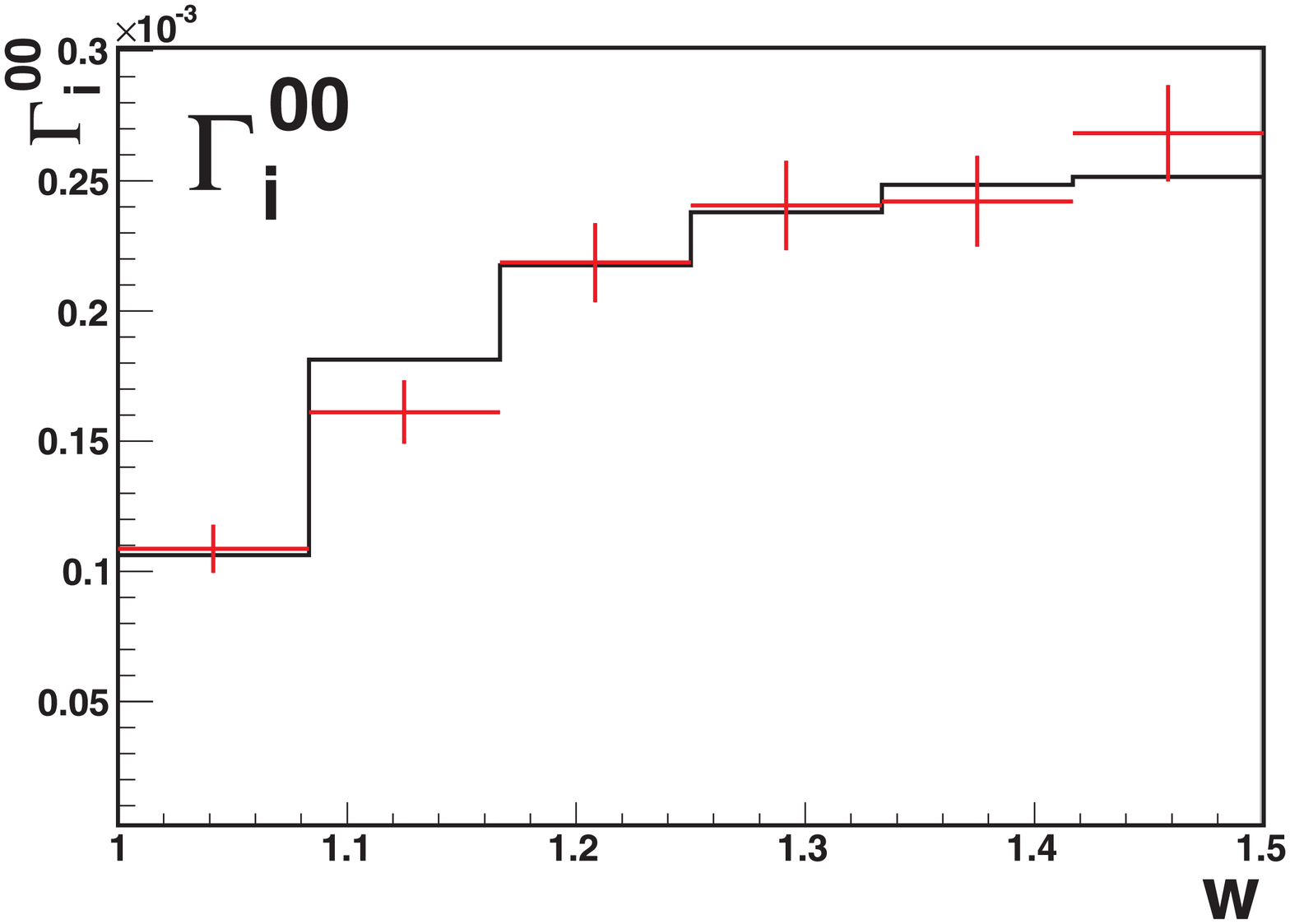}
    \includegraphics[width=0.4\columnwidth]{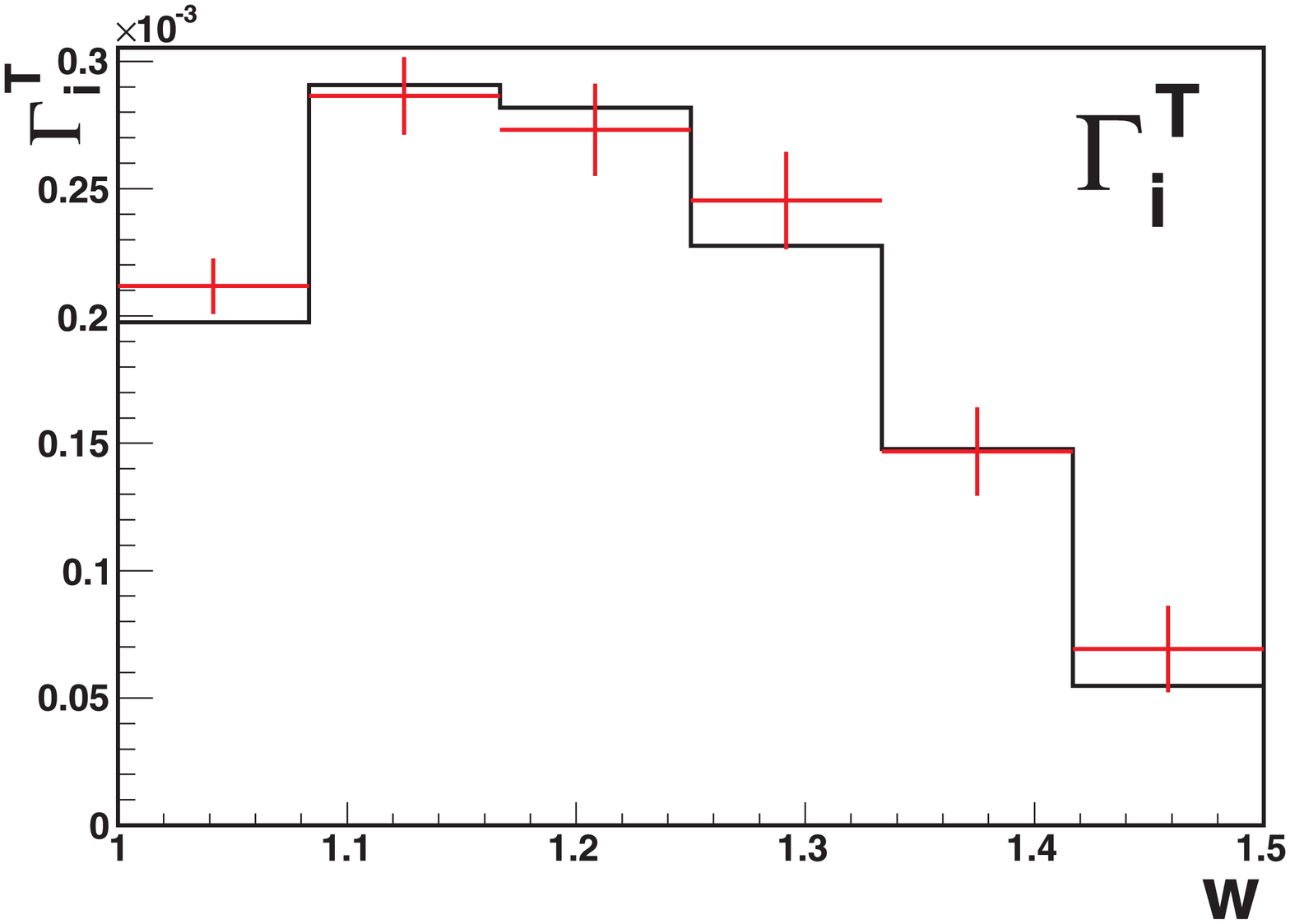}
  \end{center}
  \caption{Results of the fit of the helicity amplitudes (red crosses)
  compared to the prediction obtained by using the parametrization 
  prescription by Caprini {\it et al.}~\cite{Caprini:1997mu} (solid black line). 
  The left plot shows the results for
  $\Gamma^{00}_i$, the right one for $\Gamma^{T}_i$. Only the
  statistical error is shown. } \label{fig:8}
\end{figure}

Given the very high amount of background in the $D^0\to
K^-\pi^+\pi^-\pi^+$~mode, we use only the $K\pi, e$ and $K\pi, \mu$
channels to determine the partial decay widths for each of the helicity
components. Tables~\ref{tab:shapes-results00} and
\ref{tab:shapes-resultsT} give the results of the fits, where the
systematic errors quoted in these tables stem from the same sources as
given in the breakdown in Table~\ref{tab:6}. It is dominated by the
track reconstruction errors of the three charged tracks $\ell, \pi, K$
and the uncertainty of the $\pi^0_s$ reconstruction. The $\chi^2$ of
the fit is in good statistical agreement with the number of degrees of
freedom, we obtain $\chi^2/\mathrm{ndf}=82.6/60$ or a $\chi^2$ probability
of $P_{\chi^2}=2.8\%$. The results are shown in Fig.~\ref{fig:8}. There
is good agreement with the results of the parametrized fit.

\begin{table}
  \begin{center}
\begin{tabular}{l|@{\extracolsep{.2cm}}cc}
\hline\hline

           &  $D^0 \to K\pi. \ell = e$ &  $D^0 \to K\pi. \ell = \mu$ \\
\hline
$\Gamma^{T}$, $w \in (1, \frac{13}{12} )$ &          (     2.267  $\pm$     0.153  $\pm$     0.264  )$\times 10^{-4}$ &                      (     1.939  $\pm$     0.152  $\pm$     0.228  )$\times 10^{-4}$ \\

$\Gamma^{T}$, $w \in (\frac{13}{12}, \frac{7}{6} )$ &          (     2.695  $\pm$     0.214  $\pm$     0.307  )$\times 10^{-4}$ &                      (     3.015  $\pm$     0.216  $\pm$     0.348  )$\times 10^{-4}$ \\

$\Gamma^{T}$, $w \in (\frac{7}{6}, \frac{15}{12} )$ &          (     2.786  $\pm$     0.253  $\pm$     0.310  )$\times 10^{-4}$ &                      (     2.678  $\pm$     0.261  $\pm$     0.299  )$\times 10^{-4}$ \\

$\Gamma^{T}$, $w \in (\frac{15}{12}, \frac{8}{6} )$ &          (     2.298  $\pm$     0.249  $\pm$     0.246  )$\times 10^{-4}$ &                      (     2.673  $\pm$     0.295  $\pm$     0.290  )$\times 10^{-4}$ \\

$\Gamma^{T}$, $w \in (\frac{8}{6}, \frac{17}{12} )$ &          (     1.557  $\pm$     0.242  $\pm$     0.162  )$\times 10^{-4}$ &                      (     1.369  $\pm$     0.250  $\pm$     0.144  )$\times 10^{-4}$ \\

$\Gamma^{T}$, $w \in (\frac{17}{12}, 1.5 )$ &          (     0.588  $\pm$     0.205  $\pm$     0.056  )$\times 10^{-4}$ &                      (     0.862  $\pm$     0.284  $\pm$     0.099  )$\times 10^{-4}$ \\
\hline\hline
\end{tabular}  

\vspace{.3cm}
	
\begin{tabular}{l|@{\extracolsep{.2cm}}cc}
\hline\hline
           & fit to total sample & central value of parametrized fit \\
\hline
$\Gamma^{T}$, $w \in (1, \frac{13}{12} )$ &          (     2.117  $\pm$     0.108  $\pm$     0.248  )$\times 10^{-4}$ &1.975            $\times 10^{-4}$            \\

$\Gamma^{T}$, $w \in (\frac{13}{12}, \frac{7}{6} )$ &          (     2.865  $\pm$     0.152  $\pm$     0.327  )$\times 10^{-4}$ &2.908  $\times 10^{-4}$            \\

$\Gamma^{T}$, $w \in (\frac{7}{6}, \frac{15}{12} )$ &          (     2.732  $\pm$     0.181  $\pm$     0.303  )$\times 10^{-4}$ &2.819  $\times 10^{-4}$            \\

$\Gamma^{T}$, $w \in (\frac{15}{12}, \frac{8}{6} )$ &          (     2.454  $\pm$     0.191  $\pm$     0.263  )$\times 10^{-4}$ &2.276  $\times 10^{-4}$            \\

$\Gamma^{T}$, $w \in (\frac{8}{6}, \frac{17}{12} )$ &          (     1.468  $\pm$     0.174  $\pm$     0.154  )$\times 10^{-4}$ &1.478  $\times 10^{-4}$            \\

$\Gamma^{T}$, $w \in (\frac{17}{12}, 1.5 )$ &          (     0.693  $\pm$     0.170  $\pm$     0.070  )$\times 10^{-4}$ & 0.547  $\times 10^{-4}$            \\
\hline\hline
\end{tabular}  
  \end{center}
  \caption{Obtained results for $\Gamma_{i}^{T}$, compared to the central
    values of the parametrized fit.} 
  \label{tab:shapes-results00}
\end{table}

\begin{table}
  \begin{center}
\begin{tabular}{l|@{\extracolsep{.2cm}}cc}
\hline\hline

           &  $D^0 \to K\pi, \ell = e$ &  $D^0 \to K\pi, \ell = \mu$ \\
\hline
$\Gamma^{00}$, $w \in (1, \frac{13}{12} )$ &          (     1.025  $\pm$     0.119  $\pm$     0.120  )$\times 10^{-4}$             &          (     1.176  $\pm$     0.146  $\pm$     0.137  )$\times 10^{-4}$ \\

$\Gamma^{00}$, $w \in (\frac{13}{12}, \frac{7}{6} )$ &          (     1.544  $\pm$     0.165  $\pm$     0.176  )$\times 10^{-4}$             &          (     1.689  $\pm$     0.177  $\pm$     0.192  )$\times 10^{-4}$ \\

$\Gamma^{00}$, $w \in (\frac{7}{6}, \frac{15}{12} )$ &          (     2.238  $\pm$     0.213  $\pm$     0.237  )$\times 10^{-4}$             &          (     2.121  $\pm$     0.216  $\pm$     0.238  )$\times 10^{-4}$ \\

$\Gamma^{00}$, $w \in (\frac{15}{12}, \frac{8}{6} )$ &          (     2.677  $\pm$     0.244  $\pm$     0.268  )$\times 10^{-4}$             &          (     2.059  $\pm$     0.240  $\pm$     0.228  )$\times 10^{-4}$ \\

$\Gamma^{00}$, $w \in (\frac{8}{6}, \frac{17}{12} )$ &          (     2.406  $\pm$     0.235  $\pm$     0.256  )$\times 10^{-4}$             &          (     2.426  $\pm$     0.263  $\pm$     0.263  )$\times 10^{-4}$ \\

$\Gamma^{00}$, $w \in (\frac{17}{12}, 1.5 )$&          (     2.907  $\pm$     0.250  $\pm$     0.301  )$\times 10^{-4}$             &          (     2.384  $\pm$     0.273  $\pm$     0.278  )$\times 10^{-4}$ \\
\hline\hline
\end{tabular}  

\vspace{.3cm}
	
\begin{tabular}{l|@{\extracolsep{.2cm}}cc}
\hline\hline
           & fit to total sample & central value of parametrized fit \\
\hline           
$\Gamma^{00}$, $w \in (1, \frac{13}{12} )$ &          (     1.087  $\pm$     0.092  $\pm$     0.123  )$\times 10^{-4}$             &                 1.062  $\times 10^{-4}$            \\

$\Gamma^{00}$, $w \in (\frac{13}{12}, \frac{7}{6} )$ &          (     1.611  $\pm$     0.121  $\pm$     0.179  )$\times 10^{-4}$             &                 1.812  $\times 10^{-4}$            \\

$\Gamma^{00}$, $w \in (\frac{7}{6}, \frac{15}{12} )$ &          (     2.186  $\pm$     0.151  $\pm$     0.238  )$\times 10^{-4}$             &                2.175  $\times 10^{-4}$            \\

$\Gamma^{00}$, $w \in (\frac{15}{12}, \frac{8}{6} )$ &          (     2.406  $\pm$     0.172  $\pm$     0.262  )$\times 10^{-4}$             &                 2.379  $\times 10^{-4}$            \\

$\Gamma^{00}$, $w \in (\frac{8}{6}, \frac{17}{12} )$ &          (     2.421  $\pm$     0.175  $\pm$     0.258  )$\times 10^{-4}$             &                 2.483  $\times 10^{-4}$            \\

$\Gamma^{00}$, $w \in (\frac{17}{12}, 1.5 )$ &          (     2.683  $\pm$     0.186  $\pm$     0.298  )$\times 10^{-4}$             &                 2.514  $\times 10^{-4}$            \\
\hline\hline
\end{tabular}  
  \end{center}
  \caption{Obtained results for $\Gamma_{i}^{00}$, compared to the
    central values of the parametrized fit.}
  \label{tab:shapes-resultsT}
\end{table}

\section{Summary and discussion}

We have reconstructed about 27,000 $B^+\to\bar
D^{*0}\ell^+\nu_\ell$~decays in the 140~fb$^{-1}$ of Belle
$\Upsilon(4S)$~data. A fit to the theoretical expression for the
four-dimensional differential decay width (Eq.~\ref{eq:2_1}), assuming
the parameterization of the helicity amplitude given by Caprini {\it
  et al.}~\cite{Caprini:1997mu} yields a measurement of $|V_{cb}|$
times the form factor normalization at zero recoil,
$\mathcal{F}(1)|V_{cb}|=(35.0\pm 0.4\pm 2.2)\times 10^{-3}$. At the
same time we determine the parameters of the Caprini {\it et al.}
parameterization, $\rho^2=1.376\pm 0.074\pm 0.056$, $R_1(1)=1.620\pm
0.091\pm 0.092$, $R_2(1)=0.805\pm 0.064\pm 0.036$. The branching
fraction of the decay $B^+\to\bar D^{*0}\ell^+\nu_\ell$ is measured to
be $(4.84\pm 0.04\pm 0.56)\%$. For all numbers quoted here, the first
error is the statistical and the second is the systematic
uncertainty. All results are preliminary. These measurements are in
agreement with previous investigations of the decay $B^+\to\bar
D^{*0}\ell^+\nu_\ell$~\cite{Albrecht:1991iz,Adam:2002uw,Aubert:2007qs}.

A direct, model-independent determination of the form factor shapes
has also been carried out and shows good agreement with the HQET based
form factor parametrization by Caprini {\it et
  al.}~\cite{Caprini:1997mu}.


\begin{thebibliography}{99}
\bibitem{ref:0} Throughout this note charge conjugation is implied.
\bibitem{ref:2} Quantities evaluated in the c.m.\ frame are denoted by
  an asterisk.
\end{thebibliography}
\end{document}